\documentclass{aa}  

\usepackage{graphicx}
\usepackage{txfonts}
\usepackage{hyperref}

\begin{document}

   \title{Unveiling the 3D structure of nova shells with MUSE - The case of RR Pic}

   \author{L. Celedón
          \inst{1}
          \and
          L. Schmidtobreick\inst{2}
          \and
          C. Tappert\inst{1}
          \and
          F. Selman\inst{2}
          }

   \institute{Instituto de Física y Astronomía, Universidad de Valparaíso, Valparaíso, Chile\\
              \email{lientur.celedon@postgrado.uv.cl}
         \and
             European Southern Observatory, Santiago, Chile
             }

  \abstract
   {Nova eruptions occur in cataclysmic variables when enough material has been accreted onto the surface of the white dwarf primary. As a consequence, the material that has been accumulated until then is expelled into the interstellar medium, forming an expanding nova shell around the system. Understanding the physical process that shapes the morphology of nova shells is essential to fully comprehend how the ejection mechanism operates during nova eruptions. Because of its closeness and age, the nova shell around the classical nova RR Pic (Nova Pic 1925) is an ideal target for studying the evolving morphology of nova shells.}
   {The use of integral field spectroscopy (IFS) is a technique that has received little attention in the study of nova shells, despite the advantages in using it when studying the morphology and kinematics of nova shells. In this work, we present an IFS study of the RR Pic nova shell, with a particular emphasis on the extraction of the 3D morphology of the shell.}
   {The nova shell was observed by the Multi-Unit Spectroscopic Explorer (MUSE) instrument placed at the ESO-VLT. By measuring the extension of the nova shell in these new observations, and comparing it against previous measurements, we were able to determine the expansion history of the ejected material. We used this information, together with the distance to the system based on Gaia EDR3 parallaxes, and the systemic velocity of the system reported in the literature to obtain the physical 3D view of the shell.}
   {The MUSE datacube confirms the presence of the nova shell in H$\rm\alpha$, H$\rm\beta$ and [O{\sc iii}], and very faintly in [N{\sc ii}]. A comparison with previous observations suggests that the shell continues in its free-expansion phase but with the different parts of the shell apparently expanding at different rates.
   The data analysis corroborates the previous vision that the shell is composed of an equatorial ring and polar filaments traced by H$\rm\alpha$. At the same time, the new data also reveal that [O{\sc iii}] is confined in gaps located in the tropical regions of the shell where no Hydrogen is observed.
   The flux measurements indicate that $\sim$99\% of the shell flux is confined to the equatorial ring, while the polar filaments show a flux asymmetry between the NE and SW filaments, with the latter being $\sim$2.5 times brighter. We have estimated the mass of the shell to be $\sim$5$\times$10$^{-5}$M$_\odot$.
   From the analysis of the 3D-extracted data, we determine that the ring structure extends $\sim$8\,000 au from the central binary, and has a position angle of $\sim$155 deg and an inclination of $\sim$74 deg. The analysis of the equatorial ring reveals it is composed of a main ring and several small clouds, extending up to a height of $\sim$4\,000 au above and below the main plane of the equatorial ring. The radial profile of the whole ring structure is reminiscent of a bow shock.
   }
   {Our data have proven the capabilities of observing nova shells using IFS, and how the nova shell around RR Pic is an interesting object of study. Further and continuous observations of the shell across the electromagnetic spectrum are required to confirm the results and ideas presented in this work.}


   \maketitle
%

\section{Introduction}

Cataclysmic variables (CVs) are binary systems where a low-mass secondary star is transferring mass onto the surface of a white dwarf (WD) primary through Roche Lobe overflow. In classical CVs, the material forms an accretion disc around the WD primary before being accreted into its surface.
Because of the low-mass transfer rates, the accreted material does not burn steadily; however, instead, it accumulates onto the WD surface until the bottom layers degenerate, allowing the hydrogen to suddenly ignite, causing a thermonuclear runaway (TNR) process that lasts a couple of minutes \citep{Starrfield16}. While the energy released during the TNR is enough to increase the brightness of the system by 8 to 15 mag, it does not destroy the WD or the CV. This allows the event to continue to repeat recurrently in scales of $\sim$10\,000 years \citep{Schmidtobreick15, Hillman20}.

As a consequence of the nova event, the material that has been accreted is ejected into the interstellar medium (ISM), forming what is called a nova shell around the system. The velocities at which the material is expelled can reach hundreds to thousands of kilometres per second \citep{Aydi20}, while the ejected mass is estimated to be of the order of 10$^{-5}$ to 10$^{-4}$ M$_\odot$ \citep{Gehrz98, Sahman18, Santamaria22b}.

Hydrodynamic simulations have shown that the geometry of nova shells is determined by the physical properties of the WD, for example, with slow-rotating WDs producing more spherical shells than faster-rotating WDs. These same simulations also predict the existence of tropical rings where the density of the material is higher than in the rest of the shell, with their latitude also depending on the WD rotation \citep{Porter98}.
Another relevant factor for the geometry of the shell is the velocity at which the material is expelled after the nova. Novae with faster ejecta tend to form more spherical shells, often fragmenting into several blobs. In contrast, eruptions with slower ejecta show ellipsoidal shells which include defined structures such as equatorial and tropical rings \citep{Slavin95, Santamaria22a}.
The velocity at which the material is ejected mostly depends on the mass of the WD, with the ejecta being expelled faster for a more massive WD \citep{Gehrz98}. What this is telling us is that the study of nova shells could be used to gain insights into the WD properties.

Several nova shells have been found around CVs that experienced a nova eruption \citep[e.g.][]{GillOBrien98, DownesDuerbeck00}. In many of the young shells ($\sim$50 yrs old), we can appreciate their geometry if the shell has expanded enough for us to resolve it. The image and spectroscopic data reveal ellipsoidal and spherical geometries. An observed correlation between the shell axial ratio and the $t_3$-time value (the time in days for the nova light curve to decrease 3 magnitudes from its maximum brightness) of their progenitor nova supports the idea that faster novae tend to generate more spherical shells \citep{Santamaria22a}. 
In the case of older shells ($\sim$100 yrs), their geometry is less clear, and they are usually fragmented which makes it more difficult to discern their original geometry \citep[see for example][]{Liimets12, CastroSegura21}.

By studying nova shells at different epochs we can understand the evolution of their luminosity, velocity, and geometry. The luminosity of nova shells in H$\rm\alpha$ and [O{\sc iii}] shows an initial gentle decline with time, followed by a steeper decline until a tentative plateau is reached \citep{Downes01, Tappert20}. This behaviour was interpreted as, first, the consequence of transitioning from a thick to a thin shell (gentle decline), then the result of an expanding, optically thin shell (steeper decline),  and lastly by the shock produced during the interaction with the ISM (tentative plateau).
Studies regarding the expanding velocity evolution of nova shells present ambiguous results. \citet{Duerbeck87a} studied four shells, finding evidence for deceleration in all of them, letting him conclude that the velocity at which the nova shells expand decreases by fifty per cent approximately every 65 years. However, \citet{Santamaria20} performed a similar analysis on five nova shells \citep[including two from the sample of][]{Duerbeck87a} and did not find evidence for deceleration in any of them, suggesting that the free expansion phase in these nova shells is larger than was previously thought.
Investigating the geometric evolution of shells requires observations spanning several years. The best-studied nova shell in this regard is GK Per (Nova Per 1901) for which several images and spectroscopic data have been collected, particularly during the last decade \citep{Liimets12, Harvey16}. The optical data reveal a shell composed of several bullet-like knots of material expanding radially without evident signs of deceleration. Its overall geometry and kinematics are consistent with a cylindrical shape. Additional X-ray and radio emissions have been detected around GK Per, which have been associated with interactions between the expanding shell and the dense ISM that surround it \citep{Anupama&Kantharia05, Balman05, Takei+15}.

Most of these studies have been done using narrow-band images, long-slit spectroscopy, or a combination of both. A natural evolution of this process is the study of these objects using integral field spectroscopy (IFS). Surprisingly, the number of studies involving IFS and nova shells is very low, involving only a few particular cases. The first works were carried out for the systems V723 Cas \citep{Lyke09} and HR Del \citep{Moraes09}. In these studies the integral field unit (IFU) analysis allows the authors to conduct a spatio-kinematical analysis of the shells, revealing morphological differences for different lines within the shells in both of them.
As part of the helium nova V445 Pup analysis, IFU observations were carried out to study the kinematics of the expanding shell, revealing its bipolar nature \citep{Woudt09, Macfarlane14}.
Recently, \citet{Takeda22} and \citet{Santamaria22b} studied the young ($\sim$ 7 yrs old) shells around V5668 Sgr and QU Vul, respectively, and obtained a 3D view of the shells as part of their analysis. In both works, the 3D reconstruction reveals a compact shell, which is expected as the shells have not had time to expand enough to reveal their geometry yet.

RR Pic (Nova Pic 1925) is one of the closest post-nova systems \citep{Ramsay17}.
The nova occurred in May 1925, reaching a maximum brightness of V = 1.0 mag and $t_3$-time of 122 d \citep{Lunt1926}, for which it is usually referred to as a slow nova \citep{Duerbeck87CatalogueNovae, Strope10}. Its light curve presented a small jittering, which was particularly strong between days 60 and 70 after the eruption, and this is the reason for which it was classified as a J-type one by \citep{Strope10}.

Its nova shell was first detected 6 years after the eruption \citep{SpencerJones31} and has since been observed again between the late 1970s and the mid-1990s \citep{Williams79, Duerbeck87d, Evans92_RRPic, GillOBrien98}.
Because of the proximity of the system, it was possible to resolve its shell from very early stages.
The first spectroscopic observations carried out by \citet{SpencerJones31} reveal two bright knots at opposite sides of the remnant. Posterior observations using H$\rm\alpha$+[N{\sc ii}] narrow band filters showed that these knots have been expanding in the NE-SW direction with a position angle (PA) of $\sim$70 deg, while an additional structure appears in the SE-NW axis with a PA of $\sim$ 150 deg \citep{Williams79, Duerbeck87d}. These authors refer to these structures as the 'polar blobs or filaments' and the 'equatorial ring', respectively.
The last H$\rm\alpha$+N[{\sc ii}] narrow band image of the nova shell was obtained by \citet{GillOBrien98} (GO98 hereafter) in February 1995. Their image shows that the aforementioned structures are still discernible, with extensions of 30 and 23 arcsec for the polar blobs and the equatorial ring, respectively. The authors concluded that both structures have been expanding at a constant rate since 1931.
Spectral analysis of the nova shell in the optical and ultra-violet wavelength range has revealed a difference in the observed strength line of Carbon, Oxygen and Nitrogen between the polar blobs and the equatorial ring, with the former presenting higher C/O and O/N ratios compared with the equatorial ring \citep{Williams79, Duerbeck87d, Evans92_RRPic}.

However, since these analyses, there have been no recent studies on the shell, so its evolution over the last 25 years has remained unknown.
During these decades most of the efforts have been put into the study of the system itself, allowing us to determine its basic properties.
The light curve of the system in the optical wavelength shows shallow eclipses with a variable amplitude, which indicates an intermediate inclination of the system, estimated to be around 60 deg \citep{Schmidtobreick+2003, Sion17}.
A tomographic study of the system allowed \citep{Schmidtobreick+2003} to conclude that the eclipsing part corresponds to an emission source of the leading side of the accretion disc but also that RR Pic shares many similarities with the SW Sextantis subclass of CVs.
Its orbital period has been measured to be 3.48 hour \citep{Schmidtobreick08, Vogt+2017}. Sporadic positive superhumps associated with an eccentric accretion disc have also been detected in the system with a period of 3.79 hours \citep{Schmidtobreick08, Fuentes-Morales18}, as well as quasi-periodic oscillations around 13 minutes \citep{Kubiak1984, Schmidtobreick08}. 
The mass of the WD has been estimated to be close to the solar mass, while the secondary star should have a mass between 0.3-0.4 M$_\odot$ \citep{Haefner&Metz1982, Sion17}. Its systemic velocity has been measured to be 1.8(2) km s$^{-1}$ \citep{Ribeiro06}, with the digit in parenthesis indicating the 1$\sigma$ uncertainty in the last digit.
From its well-constrained parallax of 1.99(2) mili arcsec thanks to Gaia EDR3 \citep{GaiaEDR3}, the distance to the system can be determined to be $\sim$500 pc.
The combination of the proximity, age, and well-known properties of its CV, makes the nova shell around RR Pic an ideal candidate to test the capabilities that IFS offers, in particular the determination of a tridimensional view of the system, but also to study the possible correlations between shell and CV properties.

In this work, we present an IFS study of the nova shell around RR Pic. In Section~\ref{sec:data_acquisition} we present the details of our observations, with the analysis of the data presented in Section~\ref{sec:image_analysis}.
In Section~\ref{sec:3d_reconstruction} we explain our methodology to obtain a 3D view of the shell, as well as the results we could derive from it. The implications of our results are discussed in Section~\ref{sec:discussion} and, lastly, a summary of the work is presented in Section~\ref{sec:conclusions}.

\section{Data acquisition and quality} \label{sec:data_acquisition}

\begin{table*}
\caption{General information about RR Pic and our observations. Celestial coordinates, distance, the date of the nova, $t_3$, observation date, and exposure time are presented. Distances are based on parallaxes provided by Gaia eDR3.}
\label{tab:data}
\centering
    \resizebox{0.92\textwidth}{!}{
    \begin{tabular}{c c c c c c c c c}
        \hline \hline
            & RA (J2000)  & Dec (J2000)  & Distance[pc]   & Date of Nova  & $t_3$[d]  & Date-Obs   & MUSE mode & $t_{\mathrm{exp}}$[s] \\ \hline
        RR Pic    & 06:35:36.082 & -62:38:24.375 & 501(5)  & May 25, 1925$^{(1)}$ & 122$^{(2)}$ & Dec 12, 2021 & WFM-NOAO-N & 5600       \\
        \hline
    \end{tabular}
    }
    \tablebib{(1)~\citet{Lunt1926}, (2)~\citet{Strope10}
    }
\end{table*}

RR Pic was observed using the Multi-Unit Spectroscopic Explorer \citep[MUSE][]{Bacon10}, placed at the Very Large Telescope (VLT) from the European Southern Observatory (ESO). MUSE is an Integral Field Spectrograph covering the electromagnetic spectrum from the visual (465 nm) to the near-infrared (930 nm). When observing in its standard observation mode (WFM-NOAO-N) it provides a field of view of 1x1 arcmin$^2$ with a resolution of 0.2x0.2 arcsec$^2$ per spaxel. The instrument's capabilities make it an ideal tool for the observation of extended sources like nova shells, and its wavelength coverage will allow us to study the main lines observed in these kinds of objects.

Our observations of RR Pic were carried out as part of the so-called Apocalypse programme: a filler programme to observe targets during bad weather, non-photometric conditions, with the standard observation mode when no other observations could be made (run ID: 108.21ZY.003). A total of eight different observations of RR Pic were carried out on the night of December 12, 2021, spanning a total of 5600 seconds of observations. Because of the far from optimal weather conditions, the seeing reached values above 1.2 arcsec in all these observations. Nevertheless, these bad conditions did not diminish our capacity to study the spatio-kinematic properties of the expanding shell. The eight datacubes were reduced, flux calibrated and combined following the standard procedures using the MUSE Data Reduction Software \citep{Weilbacher20}.

A summary of our observations and some basic properties of RR Pic are presented in Table~\ref{tab:data}. The distance to RR Pic was estimated from Gaia EDR3 parallax using the Bayesian approach suggested by \citet{Luri18}. The presented distance corresponds to the mean of the posterior distribution, with the uncertainty indicating its standard deviation.

To check the quality of the MUSE datacube's flux calibration under these non-photometric conditions, we compare the magnitudes of the field stars close to RR Pic against their magnitude in the SkyMapper Southern Sky Survey catalogue \citep{SkyMapperDR1, SkyMapperDR2}. We chose the SkyMapper filters r and i as both are fully covered by the MUSE wavelength range to create bandpass images from which we can compare the MUSE photometry against the SkyMapper one and determine how much flux was lost because of the non-photometric conditions.

The SkyMapper catalogue provides photometric measurements in the AB system based on a Petrosian photometric aperture for each one of its filters. We determine the Petrosian magnitudes in the MUSE images by using Sextractor\footnote{\href{https://sextractor.readthedocs.io/en/latest/Introduction.html}{Sextractor's main page}} \citep{Sextractor} with their default parameters. The bandpass and zero point of both filters were obtained from the data provided by the SVO filter service \footnote{\href{http://svo2.cab.inta-csic.es/theory/fps/index.php?}{SVO Filter Profile Service}} \citep{SVOFilterService}.

The results of the comparison are presented in Table~\ref{tab:skymapper_comparison}. For the two brightest stars within the sample, the differences in magnitudes are less than 0.03 mag, while for the faintest the observed differences are not greater than 0.15 mag in the r filter. The comparison indicates that no major loss of flux occurs during the observations.

\begin{table*}
    \centering
    \caption{Comparison between SkyMapper and MUSE Petrosian magnitudes for the SkyMapper filters r and i. The SkyMapper object id, celestial coordinates in J2000 and SkyMapper and MUSE magnitudes are presented. The stars are sorted based on their r\_SM magnitude. Uncertainty in MUSE magnitudes comes from the uncertainty in flux given by Sextractor.}
    \label{tab:skymapper_comparison}
    \begin{tabular}{c c c c c c c}
    \hline \hline
    object\_id & RA(J2000)    & Dec(J2000) & r\_petro\_SM & r\_petro\_MUSE & i\_petro\_SM & i\_petro\_MUSE \\
    \hline
    471726387  & 6:35:40.333  & -62:38:41.738 & 14.933(6) & 14.920(1) & 14.750(8) & 14.724(1) \\
    471726565  & 6:35:36.811  & -62:38:47.529 & 16.139(8) & 16.132(1) & 15.98(2)  & 15.979(1) \\
    471726390  & 6:35:39.294  & -62:38:30.327 & 19.5(1)   & 19.647(3) & 19.46(15) & 19.395(3) \\ 
    \hline
    \end{tabular}
\end{table*}

\section{Imaging and spectral analysis} \label{sec:image_analysis}

The analysis of the MUSE datacube confirms the presence of the nova shell around RR Pic as well as the ring-like and polar filament structures previously observed. We detected the nova shell in the Balmer lines of H$\rm\alpha$ and H$\rm\beta$, as well as in the forbidden line of [O{\sc iii}] and very faintly in [N{\sc ii}].
In the following section, we present our analysis regarding the image and spectral features observed within the shell.

All the following analysis was done in {\sc Python} using the MUSE Python Data Analysis Framework package, MPDAF \footnote{\url{https://mpdaf.readthedocs.io/en/latest/}} \citep{Bacon16}.

\subsection{Image}

\begin{figure*}
\centering
\includegraphics[width=1.00\textwidth]{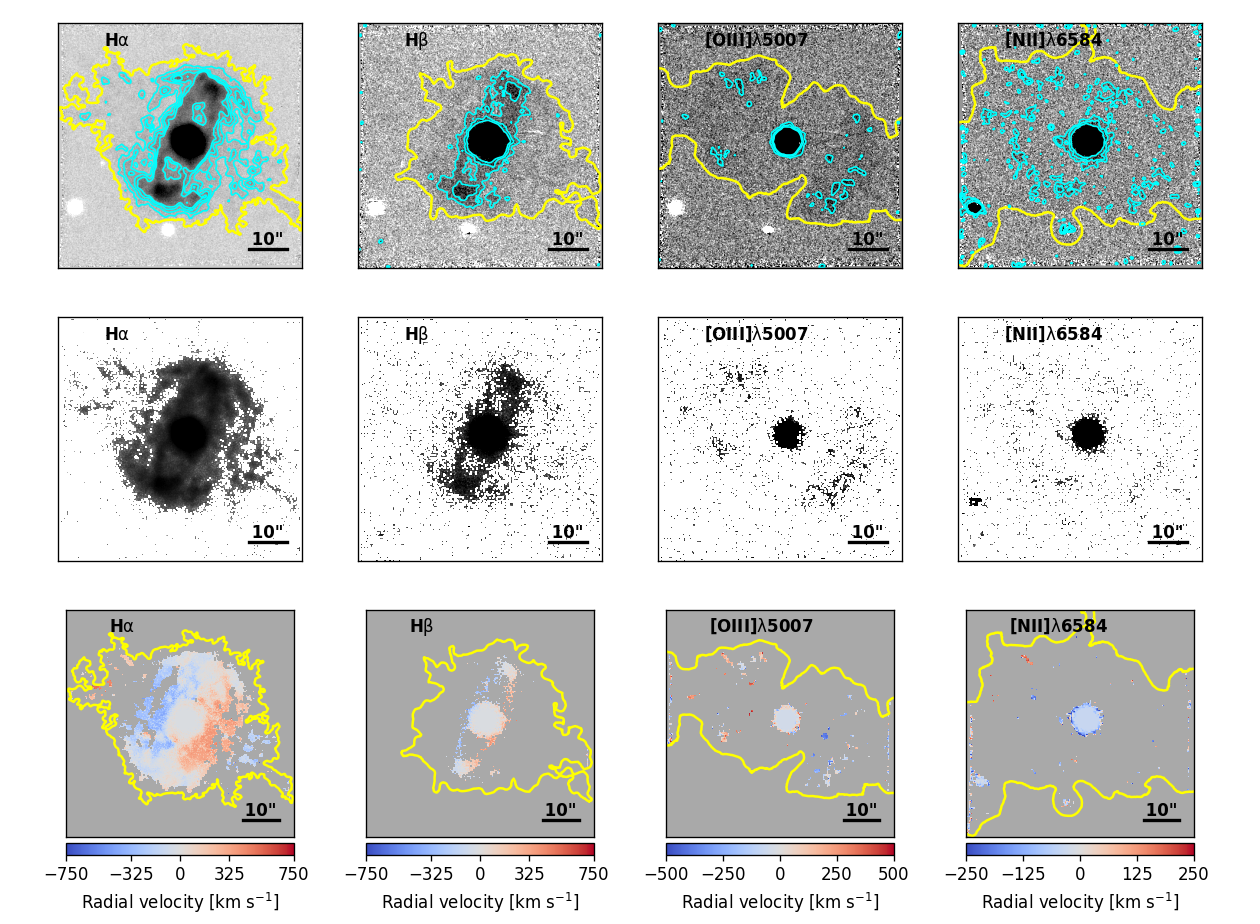}
\caption{Narrow band images and velocity maps of the RR Pic nova shell obtained from the MUSE datacube.
Top row: flux images of the shell in  H$\rm\alpha$, H$\rm\beta$, [O {\sc iii}] and [N {\sc ii}]. The yellow contour level indicates the median background value, while the cyan levels indicate fluxes up to 1,2,3 and 10 sigma over the background median for H$\rm\alpha$, 1, 2, and 3 for H$\rm\beta$ and [O{\sc iii}] and 0.5, 1, and 2 for [N{\sc ii}]. 
Middle row: same images as top row but presented in logarithmic scale to appreciate the extension of the faintest part of the nova shell. The contours presented in the top row have been removed for clarity.
The bottom row shows the radial velocity maps corresponding to each line. In all images north is up and east to the left.}
\label{fig:rr_pic_image_velocity}
\end{figure*}

The top row of Fig.~\ref{fig:rr_pic_image_velocity} presents narrow band images created from the MUSE datacube for all the detected emission lines from the nova shell. The images were created by integrating the observed emission lines within a range of velocities of [$-750$,$+750$] km s$^{-1}$ for H$\rm\alpha$ and H$\rm\beta$, [$-500$,$+500$] km s$^{-1}$ for the [O{\sc iii}] $\rm\lambda$500.7 nm line and [$-250$,$+250$] km s$^{-1}$ in the case of the [N{\sc ii}] $\rm\lambda$658.4 nm line. These velocity ranges were determined based on the observed velocity profile of the shell material so we can include all the observed emissions while minimizing the noise contribution (see Sect.~\ref{sec:spectral_description}).

These images corroborate the presence of a ring-like structure surrounding the CV which semi-major axis extends along the SE-NW direction, but also a series of polar filaments expanding along the axis defined by the SW-NE direction. For better clarity, we determine the background median value and its standard deviation through a sigma clipping process on the images after applying a 5-arcsec circular mask around the central star.
The 1 sigma contour level in the H$\rm\alpha$ image shows the equatorial ring also extending in the SW-NE direction, with several blobs of material in this direction. This overall structure composed of a clear equatorial ring and polar filaments was previously imaged by GO98, indicating that the nova shell has been expanding while preserving its geometry during the last 20 years.
We can see that these two components (ring and filaments) differ from each other in terms of their H$\rm\alpha$ flux, with the polar blobs being much fainter than the equatorial ring. For this structure, the brighter regions correspond to the zones located at each side of the ring in the SE and NW directions, which are traced by the 10-sigma contour level. The edges of the ring possess a bow shape, where it is also possible to identify regions of higher intensity. Overall, there are no significant differences between the SE and NW sections of the ring.
On the other hand, the polar filaments show a clear difference between the SW and NE regions, with the SW being brighter than its northern counterpart. We can identify several blobs of material in the southern filament interconnected between them at a 1 sigma level, while for the northern filament, we can observe four isolated blobs only.
The middle row in Fig.~\ref{fig:rr_pic_image_velocity} shows the same images but using a logarithmic scale. The extension of the faintest part of the shell can be better appreciated. Besides the already mentioned structures, an additional faint tail of material can be observed leaving the system together with the SW filaments.

The distinction between the equatorial ring and the polar filament is evident when we look at their radial velocities. The bottom row of Fig.~\ref{fig:rr_pic_image_velocity} shows the radial velocities corresponding to each one of the detected shell lines. These velocity maps reveal the NE part of the equatorial ring having negative radial velocities, indicating that this part is moving towards us, while the opposite occurs in the SW part of the equatorial ring, which shows positive velocities. This confirms the idea of an equatorial ring around the system.
In the case of the polar filaments, we have the opposite situation where the SW has negative velocities and the NE positive ones, suggesting that they are expanding in an apparent orthogonal direction with respect to the plane defined by the equatorial ring.

In the rest of the lines only one of the components, the equatorial ring or polar filaments, are present. The H$\rm\beta$ emission traces mainly the equatorial ring as it is shown in Fig.~\ref{fig:rr_pic_image_velocity}. Faint fragments of H$\rm\beta$ emission can also be observed in the logarithmic image being spatially coincident with the SW filaments observed in H$\rm\alpha$. The polar filaments are traced by H$\rm\alpha$ but also by the forbidden line of [O{\sc iii}]. In this line, we can observe several blobs at both the northern and southern filaments. Their velocity distribution shows the same pattern as the H$\rm\alpha$ blobs, with the southern blobs moving towards us. But it is also possible to observe some blobs with velocities close to zero, which indicates that they are expanding perpendicular to the observer. Lastly, the shell is also detectable in the forbidden line of [N{\sc ii}], but very faintly. The contour levels in the left, bottom row of Fig.~\ref{fig:rr_pic_image_velocity} are set to 0.5 sigmas over the median background, for which it is possible to observe several blobs of material scattered throughout the field of view, but also coincident with the locations where the equatorial ring lobes are located in H$\rm\alpha$ and with the blobs observed in [O{\sc iii}]. The brightest blob in this line is located at the NE filament, and its velocity is negative, following the behaviour of the rest of the lines.

\subsection{Comparison with the GO98 image} \label{sec:comparison_go98}

A comparison with the GO98 image reveals that the shell has expanded considerably in the last 26 years. It is expected that during the first decades after the nova event, the ejected material expands freely through the ISM, but after a certain time, it should start to show signals of deceleration, although it is not clear when this deceleration will start to become evident \citep{Duerbeck87a, Santamaria20}. To address this issue, we can compare two images taken at different times, identify features within them and compare their position in each image. This approach was used to study the expansion of the several knots that composed the nova shell around GK Per \citep{Liimets12}from which the authors conclude the knots are consistent with a free and radial expansion.
Following this scheme, we have identified several knots of material in the image published by GO98 and their tentative counterparts in the H$\rm\alpha$ MUSE image. We determined the astrometry of the GO98 image using the stars in the field as reference points, as well as RR Pic itself. This allows us to determine the celestial position of the knots to compare with the MUSE data.

\begin{figure}
\centering
\includegraphics[width=1.0\columnwidth]{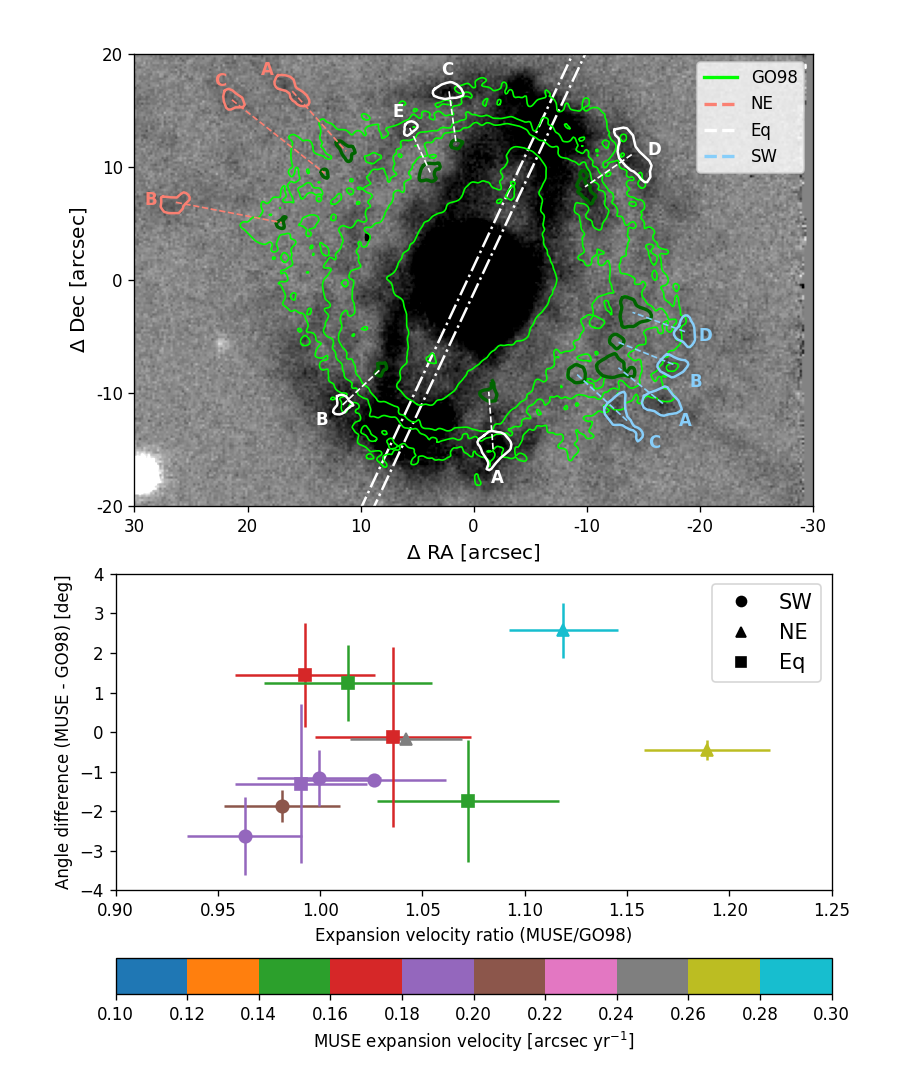}
\caption{Comparison between MUSE and GO98 H$\rm\alpha$ images. The top panel shows the MUSE background subtracted image (black-grey) with the GO98 image overimposed in green contours. Within the GO98 image, we identified several knots of material, marked with dark green contours, for which we tentatively identified their counterparts in the MUSE image (joined through dashed lines). The different colours indicate if the knots belong to the NE or SW filaments, or to the equatorial ring.
The white dotted dashed line indicates the rectangular section used to determine the extension of the equatorial ring.
The bottom panel shows the ratio of expansion velocities between the MUSE and GO98 images versus the difference in position angle between MUSE and GO98 images for each knot identified in the top panel. The different markers indicate whether the knot belongs to the equatorial ring or a filament, while the colours show the expansion velocity measured from the MUSE image. Overall, the knots are consistent with a radial, free-expanding shell.}
\label{fig:comparison_go98_muse}
\end{figure}

The top panel of Fig.~\ref{fig:comparison_go98_muse} shows the nova shell observed in H$\rm\alpha$ by MUSE, in greyscale, with the contours of the narrow-band image published by GO98 overimposed in green. Because the GO98 image does not contain physical units of flux, we arbitrarily choose the contour level values to highlight the several knots of material that we consider its geometry to have remained relatively constant in time as the shell expanded, as we were able to identify them in both, the GO98 and MUSE images.

These knots of materials are marked in dark green contours in Fig.~\ref{fig:comparison_go98_muse} when referring to the GO98 image, and in red, white or blue colours in the case of the MUSE image. The distinction in colour in the case of the MUSE image was done to differentiate the knots belonging to the equatorial ring (white) from those of the NE (red) and SW (blue) filaments. In all cases, the knots identified in the GO98 image and the MUSE data are connected through dashed lines to illustrate their expansion.

The knots are labelled from the highest surface brightness observed in the MUSE H$\rm\alpha$ image starting with A, and depending on which part of the shell they belong to. The celestial positions and surface brightness in the MUSE H$\rm\alpha$ image for each knot are presented in Table~\ref{tab:knots_info}. The positions presented for each knot correspond to the average pixel position that composes the region, weighted by their flux. We estimate the uncertainties in the position of the knots by performing this procedure 10\,000 times, each time adding a background noise following a normal distribution with $\mu$ equal to zero and $\sigma$ equal to the previously determined background standard deviation.

The bandpass used by GO98 has a width of 54 $\AA$ centred on 655.5 nm. which is broader than the wavelength range that we used to create the H$\rm\alpha$ image from the MUSE datacube (33 $\AA$ centred at 656.3 nm). Because our wavelength range is embedded within the GO98 bandpass, we do not expect significant differences in the incoming flux between both images.
The differences in bandpass transmission are irrelevant as the GO98 data was not flux-calibrated.
The seeing conditions in both images were similar, with GO98 reporting a seeing value of 1.1 arcsec during their RR Pic observations, which is comparable to the 1.2 arcsec seeing for the MUSE data. The pixel scale of the GO98 image is 0.31 arcsec per pixel, a $\sim$50\% worse than the MUSE pixel scale of 0.2 arcsec per pixel.
Because in both cases the angular resolution is dominated by the similar seeing conditions together with the similarities in bandpass, a direct comparison is possible.

For each knot in the GO98 and MUSE images, we have computed its radial distance to the central star, as well as its position angle. Then we compared the ratios in expansion velocity, which is obtained from the measured radial distance and the time since the nova, versus the difference in position angle for each knot pair. The results can be observed in the bottom panel of Fig.~\ref{fig:comparison_go98_muse}. In the case of radial free expansion of the material, and assuming we correctly identified the knots in both images, we should expect the ratio between the expansion angles to be scattered around one while the difference in position angle should be scattered around zero.
The results show that the data are overall consistent with a nova shell expanding free and radially. 

The bottom panel of Fig.~\ref{fig:comparison_go98_muse} shows that the position angle differences are scattered symmetrically around zero with differences lower than $\pm$4 degrees, consistent with a radial expansion of the shell given the uncertainties. 
But in the case of the velocity ratio we observe a systematic trend towards velocity ratios higher than one. The most evident examples correspond to the NE filaments B and C with ratios of $\sim$1.12 and $\sim$1.19 respectively.
These high-velocity ratios, could in principle indicate an acceleration of the ejected material. We find this scenario rather unlikely although not impossible, as evidence for acceleration has been reported for knots in born-again planetary nebulae \citep{Fang+2014}.
Instead, we argue in favour of a wrong identification of the knots of the NE filaments within the GO98 image due to a combination of intrinsic lower fluxes of the knots and the overall lower quality of the GO98 image compared with the MUSE image.

The colours in the markers of the bottom panel in Fig.~\ref{fig:comparison_go98_muse} indicate the expansion velocity measured from the MUSE image, whose values are presented in Table~\ref{tab:knots_info}. 
We can see that the range of expansion depends on whether the knots belong to a filament, SW or NE, or the equatorial ring. This is not surprising as previous authors already have observed the polar filaments expanding faster than the equatorial ring \citep{Williams79, GillOBrien98}.
The SW knots have a range of expansion velocities close to $\sim$0.2 arcsec yr$^{-1}$, with the exception being SW-C which shows a smaller expansion velocity of $\sim$0.18 arcsec yr$^{-1}$. In the case of the equatorial knots, the observed range of velocities is distributed approximately uniformly between $\sim$0.15 and $\sim$0.18 arcsec yr$^{-1}$. Lastly, the knots belonging to the NE filaments show the highest expansion velocities, reaching velocities between $\sim$0.24 and $\sim$0.28 arcsec yr$^{-1}$.

\begin{table*}
    \centering
    \caption{Basic information for the knots identified in the MUSE H$\rm\alpha$ image. The columns show the celestial coordinates, radial extension and position angle measured from the central star, the expansion velocity and the surface brightness for each knot. The surface brightness is given in units of 10$^{-19}$ erg s$^{-1}$ cm$^{-2}$ arcsec$^{-2}$.}
    \label{tab:knots_info}
    \begin{tabular}{ccccccc}
        \hline \hline
        Knot &   RA   &   Dec   &    $r$   &    PA    & $v_{\mathrm{exp}}$      & Surface    \\
             & (J200) & (J2000) & [arcsec] &   [deg]  &     [arcsec yr$^{-1}$]  & brightness \\
        \hline
        NE-A & 06:35:38.43(5) & -62:38:07.76(4) & 23.25(6) &  44.38(1) & 0.2407(7) & 218(2) \\
        NE-B & 06:35:39.94(5) & -62:38:17.80(3) & 27.46(6) &  76.12(4) & 0.2842(6) & 198(3) \\
        NE-C & 06:35:39.20(5) & -62:38:08.65(4) & 26.66(6) &  53.83(1) & 0.2759(7) & 188(3) \\
        Eq-A & 06:35:35.84(1) & -62:38:39.32(1) & 15.03(1) & 186.27(3) & 0.1555(1) & 787(4) \\
        Eq-B & 06:35:37.79(1) & -62:38:35.64(1) & 16.31(2) & 133.66(5) & 0.1688(2) & 649(5) \\
        Eq-C & 06:35:36.44(2) & -62:38:07.88(1) & 16.69(2) &   8.57(5) & 0.1727(2) & 721(6) \\
        Eq-D & 06:35:34.07(2) & -62:38:13.51(1) & 17.61(3) & 308.11(9) & 0.1823(3) & 476(2) \\
        Eq-E & 06:35:36.92(2) & -62:38:11.12(2) & 14.49(2) &  23.69(3) & 0.1500(2) & 450(6) \\
        SW-A & 06:35:33.69(3) & -62:38:35.39(2) & 19.81(4) & 236.24(1) & 0.2051(4) & 311(2) \\
        SW-B & 06:35:33.56(2) & -62:38:32.17(2) & 19.02(4) & 245.85(2) & 0.1969(4) & 289(3) \\
        SW-C & 06:35:34.22(2) & -62:38:36.42(3) & 17.56(5) & 226.72(2) & 0.1817(5) & 274(2) \\
        SW-D & 06:35:33.38(2) & -62:38:29.15(3) & 19.17(4) & 255.61(8) & 0.1984(4) & 271(3) \\
        \hline
    \end{tabular}
\end{table*}

\subsection{Expansion history} \label{sec:expansion_history}

\begin{figure}
\centering
\includegraphics[width=1.0\columnwidth]{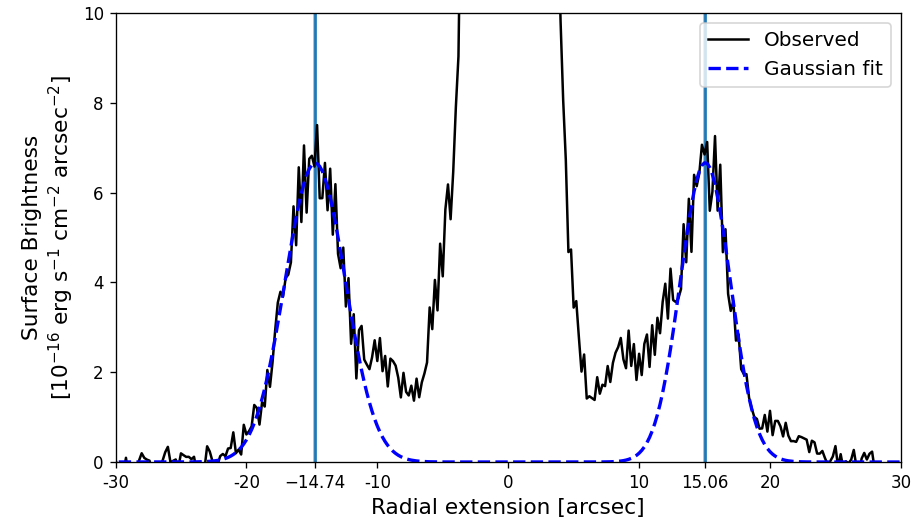}
\caption{Surface brightness profile observed in the rectangular region presented in Fig.~\ref{fig:comparison_go98_muse}. 
Gaussian profiles were fitted to each one of the observed peaks (blue dashed lines) to determine the equatorial ring extension to be 29.8(2) arcsec.
The positions of the Gaussian centres are marked with blue vertical lines.
The radial distance is expressed with respect to the central binary with positive values towards the north.}
\label{fig:equatorial_gaussian_fit}
\end{figure}

In the previous section, we have shown evidence for a still free-expanding shell around RR Pic. However, to correctly quantify the expansion history of the ejected material it is necessary to measure the shell extension in the same way as the previous authors did. They make the distinction between the equatorial ring and the polar filaments, finding that these regions expand at different velocities, similar to what we found in the previous section.
To determine the extension of the equatorial ring, previous authors measured the extension of its semi-major axis. To achieve this, we proceed then to measure the surface brightness in the H$\rm\alpha$ image using a rectangular region of 1 arcsec width and position angle of 155 degrees centred on the central binary (Top panel of Fig.~\ref{fig:comparison_go98_muse}). This region covers the semi-major axis of the equatorial ring and allows us to measure its extension as the separation between the peaks corresponding to the shell. The presence of the shell is evident as two symmetric peaks with respect to the central binary (Fig.~\ref{fig:equatorial_gaussian_fit}). We fitted a Gaussian profile to each of these peaks to estimate the separation between them based on the measurement of their centroids, giving a separation of 29.8(2) arcsec.

In the case of the polar filaments, the previous authors estimated their extension as the separation between knots, but without specifying which knots they refer to \citep{SpencerJones31, Williams79, GillOBrien98}. As we previously stated, the average radial extension of the filaments is different for the NE and SW filaments, with the first one being more extended than its southern counterpart. This means we could obtain an artificial acceleration or deceleration of the shell depending on which filament we choose to compare with the previous measurements.
In the GO98 image, the authors reported an extension of the polar filaments to be 30 arcsec. Analysing their image, we found that this value agrees well with the average separation between the NE and SW knots. This means that we can use the average radial distance between the NE and SW knots in the MUSE image to compare with the previously measured one.

\begin{figure}
\centering
\includegraphics[width=1.0\columnwidth]{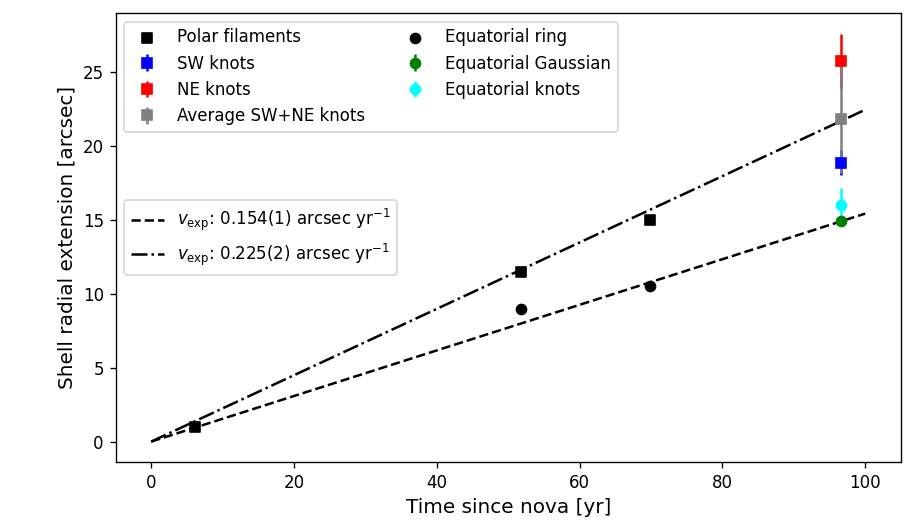}
\caption{Radial extension of the nova shell around RR Pic as a function of the years since the nova eruption. Measurements provided by previous authors are presented in black dots for the equatorial ring and squares for the polar filaments. The new measurements from MUSE data are presented in colours. The best linear model indicates expansion rates of 0.154(1) and 0.225(2) arcsec yr$^{-1}$ for the equatorial ring and the polar filaments respectively.}
\label{fig:rr_pic_shell_expansion}
\end{figure}

The radial extension of the shell around RR Pic as a function of the years since the nova is shown in Fig.~\ref{fig:rr_pic_shell_expansion}.
The shell extension measurements reported by the different previous authors are presented in black dots for the equatorial ring and black squares for the polar filaments.
The first measurement was provided by \citet{SpencerJones31} at a time when the shell structures were not resolvable. Because we currently know the polar filaments are expanding faster, we will consider this point to belong to the mentioned filaments. In the case of the other two authors, \citep{Williams79} and GO98, they could resolve both structures and therefore provide measurements for each one.

The new data are presented in colour squares for the polar filaments (red for the NE knots, blue for the SW knots, and grey for the average between the polar knots), while the equatorial ring measurements are presented in cyan and green dots for the slit measurement and the equatorial knots respectively. In the case of the measurement from knots, the markers indicate the mean value, while the error bars correspond to the standard deviation of the sample.
With the addition of these new data, we proceed to perform a linear fit to find the expansion velocity rate for the polar and equatorial ejecta. The results indicate that the equatorial ring is expanding at a rate of 0.155(1) arcsec yr$^{-1}$ (dashed line), while the polar filaments are expanding at a rate of 0.224(2) arcsec yr$^{-1}$ (dotted line). In the case of the polar filaments, we included only the average value between the NE and SW knots. These expansion velocities are consistent with the previous expansion values of 0.159(1) and 0.217(3) arcsec yr$^{-1}$ for the equatorial and polar filaments, clearly supporting the free-expanding scenario proposed in the previous section. By using the Gaia distance (Table~\ref{tab:data}), we can convert the new measurements for the projected expansion velocity from arcsec yr$^{-1}$ into proper velocity values. This led to an expansion velocity of 368(4) km s$^{-1}$ for the equatorial ring and 532(8) km s$^{-1}$ for the polar filaments.
These expansion velocities are consistent when compared with the observed spectral profile (Fig.~\ref{fig:spec}) for the case of the equatorial ring, but appear to be higher in the case of the polar filaments, as the spectra show velocities around $\sim$400 km s$^{-1}$. This is caused by the fact that the filaments are expanding orthogonally with respect to us.
If we instead consider the NE and SW radial extension measured from the knots as independent each one, then we found an average expansion of 0.199(5) arcsec yr$^{-1}$ for the SW and 0.26(1) for the NE, which translates into an expansion velocity of 473(12) and 613(26) km s$^{-1}$ respectively.

\subsection{Description of the spectrum} \label{sec:spectral_description}

For the spectral analysis of the nova shell, we make a distinction between the equatorial ring and polar filaments based on the contour levels and velocity maps presented in Fig.~\ref{fig:rr_pic_image_velocity}.
For the equatorial ring, we created a mask that considers the data within the 1 sigma level presented in the H$\rm\alpha$ image. We did not consider the filaments at the north and south as their velocities differ from the ring behaviour. Once this mask is defined, we can apply it to the datacube, allowing us to obtain a single spectrum by summing all the flux contributions at each wavelength.
To minimize the contamination from the stellar component, we defined a circular mask centred on the position of RR Pic, and with a radius of 5 arcsec. We chose this value for the mask because it offers a good equilibrium between removing the stellar flux and keeping most of the nova shell flux intact. Following the same procedure of applying this circular mask to the cube we obtained a single spectrum for the CV, which then can be subtracted from the equatorial spectrum to end up with a spectrum that represents mostly the contribution of the nova shell.
In the case of the polar filaments, we consider both filaments observed in H$\rm\alpha$ as well as the ones observed in [O{\sc iii}] at 1 sigma level with respect to the background to create a mask which yields a spectrum in the same manner as the previous case.

\begin{figure*}
    \centering
    \includegraphics[width=1.0\textwidth]{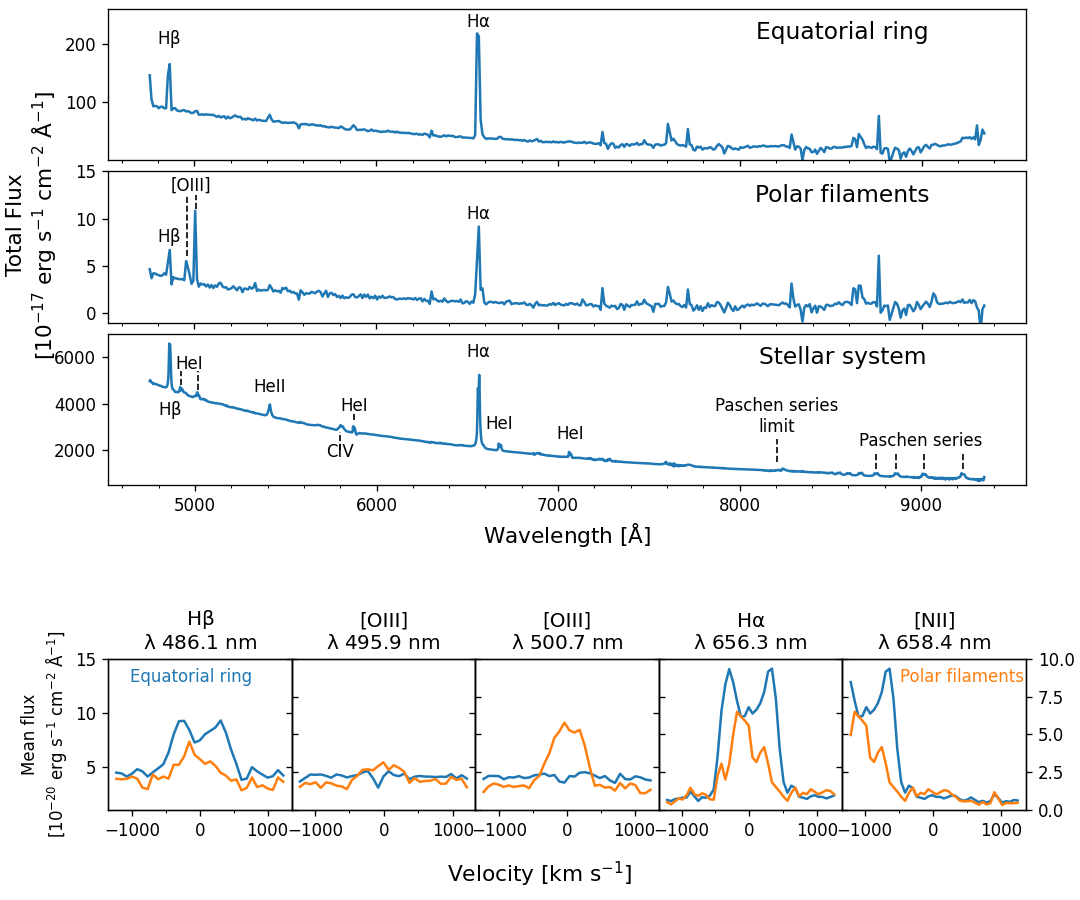}
    \caption{
    Spectra of the equatorial ring, polar filaments and the central binary (CV) of RR Pic.
    The top panels present the total spectrum of the equatorial ring, the polar filaments and the CV, with the observed main lines labelled in the spectra.
    The bottom panels show the average spectra of the equatorial ring (blue) and polar filaments (orange) for the lines observed within a velocity range between -1250 and 1250 km s$^{-1}$. The flux scales are different, with the equatorial ring defined by the scale present on the left, and the polar filaments by the scale presented on the right.
    }
    \label{fig:spec}
\end{figure*}

The resulting spectra are presented in Fig.~\ref{fig:spec}. The upper panels show the spectrum of the equatorial ring, the polar filaments and the CV itself. The total flux inside the masks is presented. The spectra of the equatorial ring and polar filaments presented in top panels were median binned by considering boxes of 8 pixels to reduce the noise within the data, while the spectrum of the CV was median binned using boxes of 2 pixels.
At each panel, the lines detected are labelled. We can see that the equatorial ring is dominated by Balmer lines, while the polar filaments include both, the Balmer lines and the [O{\sc iii}] $\rm\lambda$495.9 and $\rm\lambda$500.7 nm lines.
The stellar spectrum shows a blue continuum dominated by the accretion disc, with several emission lines that come from it. These emission lines are dominated by the Balmer lines, but also several Helium lines are observable, including the He{\sc i} $\rm\lambda$492.1 nm, $\rm\lambda$501.5 nm, $\rm\lambda$587.5 nm, $\rm\lambda$667.8 nm and $\rm\lambda$702.1 nm lines, and the He{\sc ii} $\rm\lambda$541.1 nm line. The C{\sc iv} $\rm\lambda$580.0 nm line is also prominent, as well as the Paschen series. No evident absorption lines are detected.
Additionally, in the equatorial and polar filament spectra, a blue continuum is also present. This is the result of the residual contribution of the CV flux into the background. Similarly, several features are observed beyond the 700 nm in these two spectra. An examination of the datacube reveals that they are caused by the high background noise at these wavelengths.

The bottom panels of Fig.~\ref{fig:spec} present a close-up of the main, unbinned lines detected in the nova shell, in blue for the equatorial ring and orange for the polar filaments. The mean flux instead of the total one is presented to facilitate the comparison. All lines are presented in their respective velocity space with a range between -1250 and 1250 km s$^{-1}$. The Balmer lines in the equatorial ring present a double peak profile which indicates the symmetry of the equatorial ring. In the case of the polar filaments, a more clumpy structure is observed with one central peak and two minor ones at each side. The peak also appears to be blueshifted. This is consistent with the SW filaments having negative velocities and higher flux than their NE counterpart, as also indicated in the images (Fig.~\ref{fig:rr_pic_image_velocity}).
In the case of the forbidden lines of Oxygen, we only see one clumpy peak. A small bump is present in the polar filament at the [N{\sc ii}] line, but it is difficult to characterize it due to the low signal.

\section{3D reconstruction} \label{sec:3d_reconstruction}

Our main goal is to obtain a 3D view of the shell from the data, from which we can later study its geometry, luminosity, spatial extension, and other physical properties. In the following, we explain our methodology to obtain the 3D view from the datacube, as well as the assumptions and considerations we made to achieve our goal.

\subsection{Extraction methodology}
\subsubsection{Shell extraction} \label{sec:shell_extraction}

We start the extraction of the shell structure from the datacube by discerning between what is signal and what is noise. First, we select a suitable range of spatial axis and wavelength within which we expect all the shell flux from a given line to be contained. These ranges are used to create a subcube from the original datacube.
By linearly interpolating the average of the three first and last svoxels at each spaxel of this subcube, a continuum cube is obtained, which is then subtracted from the subcube data.

With the background subtracted we start to proceed to select those svoxels that are likely to belong to the nova shell. 
As we show in Fig.~\ref{fig:rr_pic_image_velocity} and ~\ref{fig:spec}, the strongest lines regarding the nova shell are the Hydrogen $\rm\alpha$ and $\rm\beta$ lines, as well as the forbidden line of [O{\sc iii}].
We define two selection criteria to select the flux coming from these lines while avoiding the noise and artefacts present in the data.
The criterion selection is based on a minimum signal-to-noise ratio (S/N): we select only those svoxels whose fluxes are greater than a certain threshold.

The S/N can be calculated at each svoxel from the variance cube provided by MUSE according to $\mathrm{S/N} = \frac{F}{\sqrt{\sigma^2}}$, where $F$ and $\sigma^2$ are the flux and variance respectively. A low threshold value will ensure that most of the flux coming from the shell will be considered but with the detriment of including a certain portion of noise.

That is where our second criterion comes into place. To get rid of most of the noise, a minimum neighbours number criteria was applied to all those svoxels that fulfil the first criteria. Only those that have a minimum certain number of adjacent svoxels to them are selected. In that way are sure to have removed most of the pure random noise.
Lastly, we discard a certain percentage of the svoxels that have passed the second criteria selection, based on their flux values, to be sure we dispose of most if not all, the noise in the subcube.
For the case of RR Pic, we set the values for these criteria to 2 for the minimum S/N value, having at least 3 neighbouring svoxels, and the discard of the 10 per cent of the faintest svoxels.

If necessary, the S/N can be increased prior to the application of these criteria by binning the subcube. This was the case when we extracted the line of [O{\sc iii}], in which case we spatially (ra/dec axis) binned the subcube by a factor of two.

\subsubsection{Position-position-velocity space} \label{sec:ppv_space}
At this point, we have selected all those svoxels with a high flux contribution with respect to the continuum, which includes the contribution of both the nova shell and the CV, which itself is dominated by the emission coming from the accretion disc around the WD (Fig.~\ref{fig:spec}). We know the position of each svoxel and the pixel scale of the instrument, so we can determine the distance in arcsec from the central binary, as well as the velocity with respect to the reference line for each svoxel, which can be obtained from the classic Doppler effect. With that information, we can create the Position-Position-Velocity (PPV) space for each emission line in the shell, which can be used later to construct the physical 3D space.

As we are interested only in the nova shell, not the accretion disc, we proceed to remove its contribution from the PPV space. 
For that purpose, it is sufficient to remove the CV contribution by using a circular aperture.
If the shell has expanded enough to be spatially resolved from the central binary then the mask should not present any problem for further analysis. But if this is not the case and it is not possible to spatially resolve the shell from the central binary, as could be the case for younger and/or distant nova shells, then masking it would be harmful, and therefore not recommended. In these cases, the analysis should go on with the inclusion of both sources of emission.
For the case of RR Pic, we fall into an intermediate point where the shell has expanded enough to be clearly distinguished from the central star, except for the most inner parts of the equatorial ring, where, because of the inclination angle of the ring, some parts appear to be projected very close to the central star. We proceed to mask the central star using a 5 arcsec aperture which removes almost all its flux, without carving too much of the equatorial ring.

\subsubsection{Conversion to physical space} \label{sec:conversion_to_ppp}

\begin{figure}
\centering
\includegraphics[width=1.0\columnwidth]{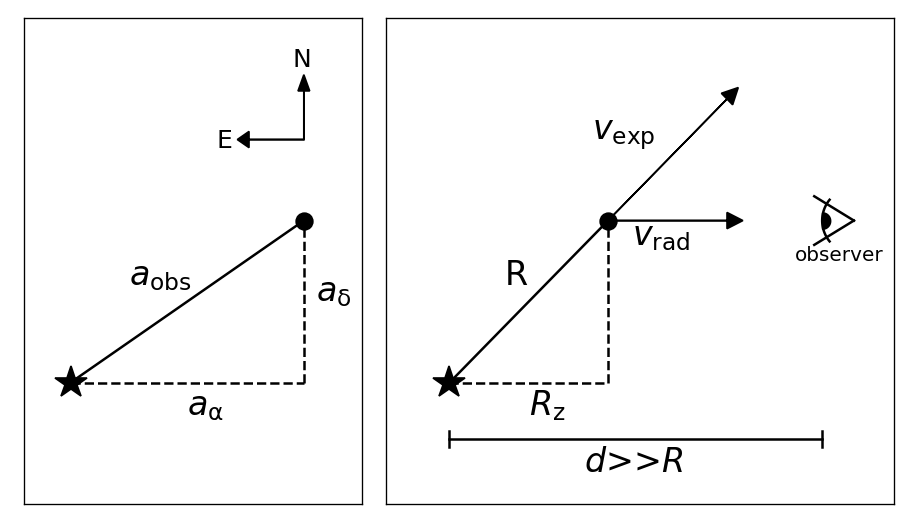}
\caption{Schematic visualization of the conversion of a single svoxel from PPV to PPP space. The left panel shows the situation in the plane of the sky, while the right panel presents the view perpendicular to it, i.e., in the line of sight of the observer. The central binary is marked with a star and the svoxel with a filled circle in both cases. See text for details.}
\label{fig:ppp_diagram}
\end{figure}

After the shell has been extracted, the PPV space can be converted into a properly physical 3D space, also called Position-Position-Position (PPP) space, but to achieve that we require some additional information. The position of a single svoxel in the plane of the sky (left) with respect to the position of the central binary is shown in Fig.~\ref{fig:ppp_diagram}.

This svoxel is located at a projected distance $a_{\mathrm{obs}}$ from the central binary. We can decompose this distance into its right ascension ($a_{\mathrm{\alpha}}$) and declination ($a_{\mathrm{\delta}}$) components. By knowing the distance from the observer to the binary, $d$, we can convert these projected coordinate differences into physical distances, $R_\mathrm{x}$ and $R_\mathrm{y}$, according to:

\begin{equation}
    R_\mathrm{x} = a_\mathrm{\alpha}\,d \quad\text{,}\quad R_\mathrm{y} = a_\mathrm{\delta}\,d .\
    \label{eqn:r_xy}
\end{equation}

Here we are assuming that the physical separation between the nova shell and the central binary is negligible compared to the distance between the central binary and the observer. Considering that the radial sizes of nova shells are of the order of $\sim$ $10^3$ AU, while the distance between the observer and post novae typically amounts to $10^2$-$10^3$ parsec, this is a very reasonable assumption.

In addition to the position of each svoxel, we also have its radial velocity. The right panel of Fig.~\ref{fig:ppp_diagram} shows the position of this svoxel on the z (line of sight) axis. The svoxel is located at a physical distance $R$ from the central star and is moving away from it at an expanding velocity $v_\mathrm{exp}$. We can only measure the projected velocity of the expanding material, $v_\mathrm{rad}$, from which we must find the distance projected on the z-axis, $R_\mathrm{z}$. If we assume that the material is expanding radially from the central star, then the angle formed by $v_\mathrm{exp}$ and $v_\mathrm{obs}$ is identical to the one formed by $R$ and $R_\mathrm{z}$, and thus easy to see the given relation: 

\begin{equation}
    R_{\mathrm{z}} = \frac{R}{v_\mathrm{exp}}\,v_\mathrm{rad} .\
    \label{eqn:r_z}
\end{equation}

This assumption does not necessarily imply a spherical geometry for the nova shell, as different sections can expand at different velocities, resulting in oblates, prolates, or more complex geometries.

To determine the quantities $R$ and $v_{\mathrm{exp}}$ we need to assume a function for the expanding history of the ejected material, which we denote as $f_{\mathrm{vexp}}(t)$. The expressions for $R$ and $v_{\mathrm{exp}}$ are then given by:

\begin{equation}
    R = \int_0^{t_{\mathrm{sn}}} f_{\mathrm{vexp}} (t) dt
    \label{eqn:R}
\end{equation}

\begin{equation}
    v_{\mathrm{exp}} = f_{\mathrm{vexp}} (t_{\mathrm{sn}})
    \label{eqn:v_tsn}
\end{equation}

where $t_{\mathrm{sn}}$ corresponds to the time that has passed since the nova event. The factor $R/v_\mathrm{exp}$ in Eq.(~\ref{eqn:r_z}) has dimensions of time, and it acts like a correction factor for the current, observed value of $v_\mathrm{rad}$.
In the case of a constant expansion velocity, which is the expected case for young shells where the material is still in its free expansion phase, this correction factor becomes simply $t_{\mathrm{sn}}$ (linear expansion with time), and the value of $R_\mathrm{z}$ is obtaining just by multiplying $v_\mathrm{rad}$ by the time since nova.
In the case of older shells where deceleration starts to become evident, a function that represents this deceleration must be assumed, and the quantities for $R$ and $v_{\mathrm{exp}}$ have to be computed accordingly. In this case, the factor $R/v_\mathrm{exp}$ will be a function of $t_\mathrm{sn}$ having units of time, but multiplied by a factor greater than one, compensating that way the underestimation in $R_\mathrm{z}$ if we would instead assume a linear expansion of the shell. Because different regions of the shell could expand and decelerate at different rates, it is important that these differences are represented correctly within our function $f_\mathrm{vexp} (t)$, otherwise, we could be assuming a spherical expanding shell which is usually not the case in nova shells.

Lastly, it is important to mention that it is essential to correctly identify the transition corresponding to the observed emission. A misidentification would lead to an erroneous radial velocity and thus to improper values for $R_{\mathrm{z}}$, which will negatively affect the recovered morphology of the shell.
While this may seem trivial, strong blends are likely to occur for the lines of H$\rm\alpha$ and [N{\sc ii}], and can pose a serious problem for the recovery of the shell structure \citep[e.g.][]{Tappert23}.
In our case the contribution of the [N{\sc ii}] line is negligible compared to the H$\rm\alpha$ emission, as it was shown in Figures ~\ref{fig:rr_pic_image_velocity} and ~\ref{fig:spec}, therefore we do not need to worry about possible blends between these two lines.

Last, but not least, the radial velocities will also have to be corrected for the systemic velocity $v_\mathrm{sys}$ of the central binary.

\subsection{RR Pic 3D reconstruction}

\begin{figure*}[!h]
\centering
\includegraphics[width=0.95\textwidth]{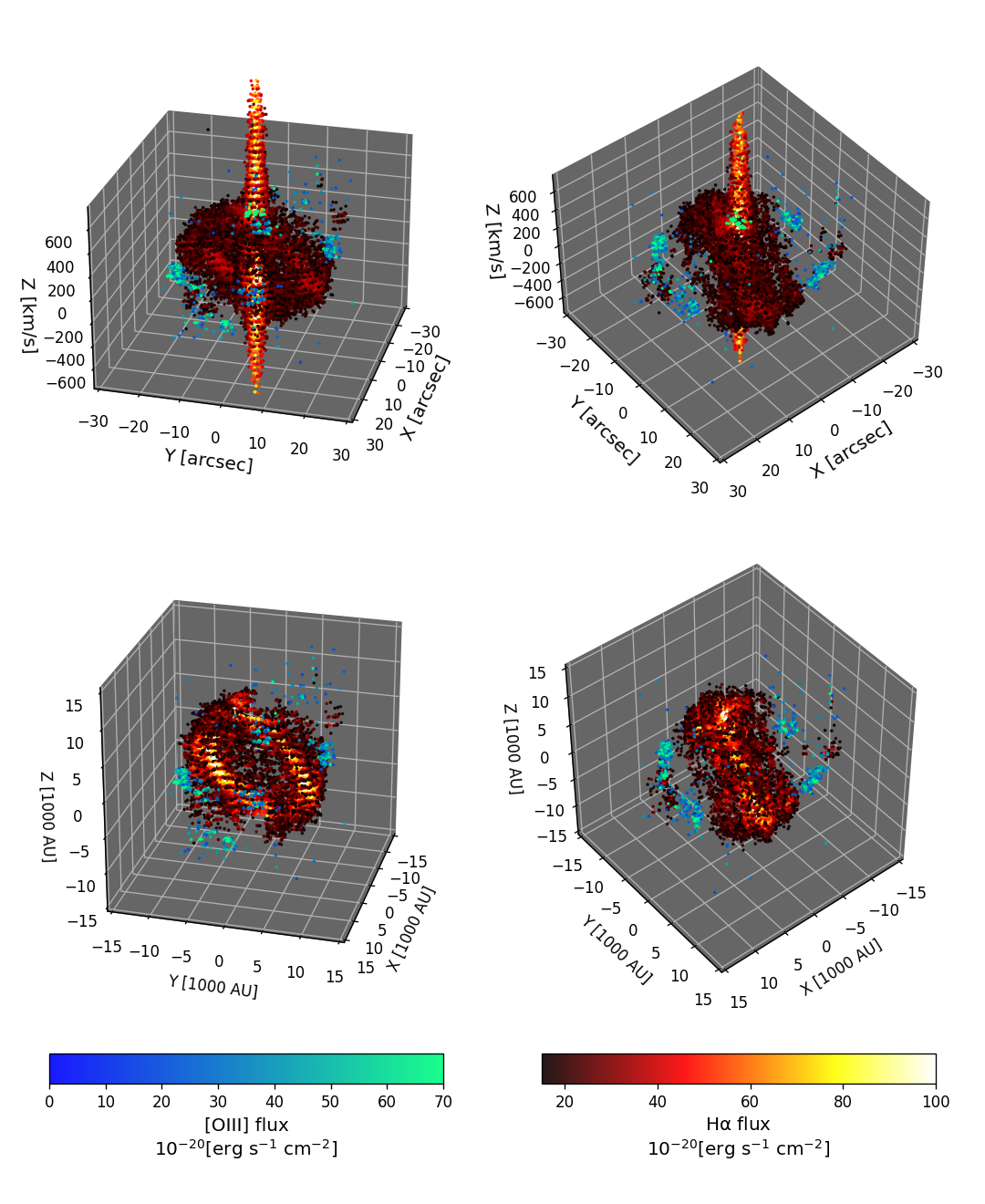}
\caption{3D view of the PPV (top row) and PPP (bottom row) spaces for the RR Pic nova shell. H$\rm\alpha$ is presented in black-red-yellow colours and [O{\sc iii}] is presented in blue-green colours. In both cases, the brighter regions are presented in lighter colours.
The angles of the frames were chosen so the circular ring (left column) and the polar filaments (right column) could be better appreciated.}
\label{fig:rr_pic_ppv_ppp}
\end{figure*}

After the extraction of the shell using the methodology we described in Section~\ref{sec:shell_extraction}, we can proceed to study the nova shell in the PPV space. The top row of Fig.~\ref{fig:rr_pic_ppv_ppp} shows a 3D view of the PPV space of the nova shell centred on the CV before the application of the circular mask explained in Section~\ref{sec:ppv_space}, where H$\rm\alpha$ and [O{\sc iii}] emissions are presented in black-red-yellow and blue-green colours respectively. The axes on the plane of the sky (X and Y) correspond to the observed projected separation in right ascension and declination respectively, and the axis along the line-of-sight (Z) to the observed projected velocity.

The first thing we can observe in the PPV space is the presence of a biconical structure appreciable in H$\rm\alpha$. This structure is the result of the high velocities reached by the material in the accretion disc around the WD and shows why its removal is important to properly analyse the nova shell.
The shell itself appears as a ring structure surrounding the system, with additional cloudy features observable above and below the ring. The aforementioned filaments are also detectable in the PPV space as clouds of material located perpendicular to the central ring. Similar structures are also evident in [O{\sc iii}], where they appear to occupy a different region in the space than the H$\rm\alpha$ emitters.

When converting to PPP after masking the central star, we considered a free-expanding model for the shell based on our results presented in Section~\ref{sec:expansion_history}. Together with the Gaia distance of 501 pc (Table~\ref{tab:data}) and the reported systemic velocity of 1.8 km s$^{-1}$, we can convert the observed quantities into proper physical distances with confidence.
The resulting 3D view of the nova shell in the PPP space is presented on the bottom row of Fig.~\ref{fig:rr_pic_ppv_ppp}. The ring structure is much clearer now after removing the contribution of the accretion disc. The position of the polar filaments in H$\rm\alpha$ and [O{\sc iii}] is more clear, with the two emitters being spatially separated, confirming the impression from the PPV image.

\subsection{Measurements from 3D data}

\subsubsection{Geometrical measurements} \label{sec:3d_geometrical_measurements}

\begin{figure}
\centering
\includegraphics[width=1.0\columnwidth]{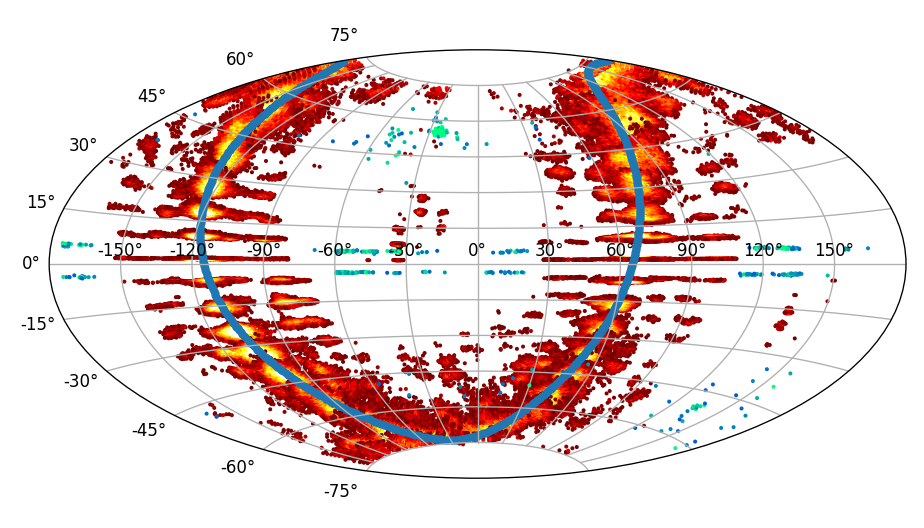}
\caption{
Aitoff projection of the RR Pic nova shell. The H$\rm\alpha$ and [O{\sc iii}] emission are presented together with the best circular ring fitted to the observed equatorial ring shown in pale blue. The colour scale for H$\rm\alpha$ and [O{\sc iii}] is the same as in Fig.~\ref{fig:rr_pic_ppv_ppp}.
}
\label{fig:rr_pic_aitoff}
\end{figure}

The creation of a 3D frame for the nova shell allows us to study its geometry in great detail. As the most evident structure within the shell is the equatorial ring, we start our analysis by fitting a circular ring to it. This will give us information about its radial size, inclination, and position angle.
We select only the 5\% of the brightest points belonging to the equatorial ring to trace the inner part of the ring but also to reduce the computational cost of the process.
The fit was done via 3D space minimization of the separation between the svoxels belonging to the equatorial ring in the PPP space and the circular ring itself. Uncertainties for the distance and systemic velocity were incorporated through a Montecarlo process by performing a thousand different fits, each one incorporating a distance and a systemic velocity value drawn from a normal distribution with mean and standard deviation being the value and its uncertainty for each one of the two aforementioned parameters. Each random sample was used to build the PPP space from the PPV as was described in Section~\ref{sec:conversion_to_ppp}, after which the best parameters to fit a circular ring to the created PPP were determined. 
The results of these thousand fits are presented in Table~\ref{tab:3d_results}, where the given values and uncertainties correspond to the mean value and standard deviation for each one of the parameters. Our results constrain very well all the ring parameters, up to the point that one could consider the given uncertainties as unrealistically low. The reason for this is we incorporated only the 5\% brightest points. If we increase the number of points within the fit, then the uncertainty values likely will also increase.
The Aitoff projection of the nova shell is presented in Fig.~\ref{fig:rr_pic_aitoff} after transforming the Cartesian coordinates of the 3D data into spherical coordinates. The H$\rm\alpha$ and [O{\sc iii}] emissions are presented in black-red and cyan-blue colours respectively.
The best fit for the circular ring is presented in blue points. It is clear that the circular ring traces the observed distribution of the equatorial ring in H$\rm\alpha$ very well, which confirms the quality of the fit for the inclination and position angles. In addition to the Hydrogen line, we can observe the distribution of the [O{\sc iii}] line, which appears concentrated around the equator of the projection ($v_{\mathrm{obs}}$ $\sim$ zero) and the polar regions, emphasizing once more the difference in spatial location between the two emitters.

\begin{table}
\centering
\caption{
Best parameters from the circular fit to the equatorial ring. Its radial extension, position angle, and inclination are presented.
}
\label{tab:3d_results}
\begin{tabular}{c c} 
\hline\hline  
Parameter & Value \\
\hline                       
Radial extension [AU] & 7890$\pm$60 \\
Position angle [°] & 155.0$\pm$0.2 \\
Inclination [°] & 73.6$\pm$0.3 \\
\hline                                  
\end{tabular}
\end{table}

\begin{figure}
\centering
\includegraphics[width=1.0\columnwidth]{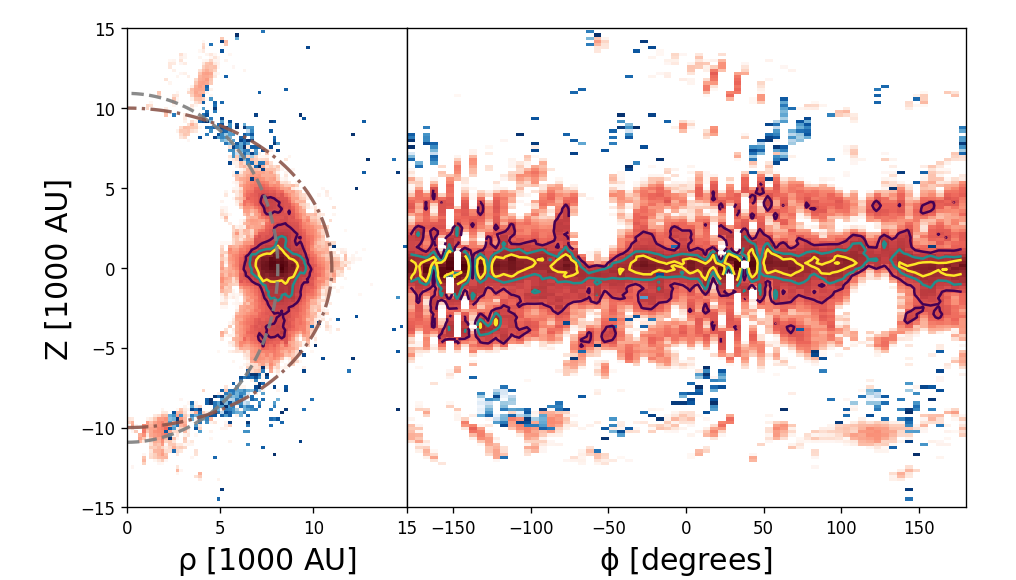}
\caption{Projection of the nova shell in cylindrical coordinates. The system is presented so the equatorial plane is located at Z=0. Left: average view on the $\rm\rho$-Z plane. Right: average view on the $\rm\phi$-Z plane. In both panels, H$\rm\alpha$ emission is presented in red-black colour scale, while [O{\sc iii}] is presented in blue. The contour levels indicate the flux corresponding to 1,2, and 3 times the standard deviation value on each plane. The lines in the $\rm\rho$-Z plane indicate an ellipse with an equatorial radius of 8\,090 au and an axial ratio of 1.35 (grey dashed line), and an equatorial radius of 10\,990 au and an axial ratio of 0.91 (brown dotted line).}
\label{fig:rr_pic_cylindrical}
\end{figure}

Once we have determined the best circular ring, we can use it to define a new reference frame in cylindrical coordinates where the circular ring, and therefore the equatorial ring of the shell, lies on the $\rm\rho$-$\rm\phi$ plane. This coordinate system allows us to better appreciate the radial structure of the equatorial ring, its overall size, and the polar filaments. In the left panel of Fig.~\ref{fig:rr_pic_cylindrical} we show the mean radial profile of the shell, that is, the $\rm\rho$-Z plane. The equatorial ring extends up to $\rho$$\sim$10\,000 AU, a value that is higher than our measurement of $\sim$8\,000 au (Table~\ref{tab:3d_results}). The reason is that the latter traces the inner and denser parts of the ring, and not the outer ones.
The equatorial ring shows a bow shock-like structure, with material above and below the main ring. To better highlight this structure in the image we provided contour levels corresponding to 1, 2, and 3 sigmas over the H$\rm\alpha$ mean value in the plane. The cloudy structures, traced at the 1$\sigma$ level, appear as regions of higher flux around 3\,900 au above and below the equatorial ring, and an opening angle of $\sim$30 deg with respect to the equatorial plane.
It is also interesting to note that the [O{\sc iii}] emission is located in the gaps between the equatorial ring and the polar filaments observed in H$\rm\alpha$. The axial ratio between both emitters is also different. The lines plotted in the $\rho$-Z plane correspond to the ellipses that best fit the observed data in H$\rm\alpha$ (grey dotted line) and [O{\sc iii}] (brown dashed line). The ellipses parameters (axial ratio and equatorial radius) were determined via least squares minimization from the data presented in the plane, weighted by their fluxes. Both, H$\rm\alpha$ and [O{\sc iii}] were fitted independently from each other. The results indicate a prolate ellipse for H$\rm\alpha$ with an equatorial radius of 8\,090(20) au and axial ratio (defined as the polar over the equatorial radius) of 1.34(3), while for [O{\sc iii}] the best fit indicates an oblate geometry with an equatorial radius of 11\,000(30) au and axial ratio 0.92(3).
In the case of the H$\rm\alpha$ data, the parameters found are in reasonable agreement with the ones determined through the 
equatorial ring fit (equatorial radius 7.89(6) versus 8.09(2)) and the ratio between the equatorial and polar expansion rates (1.445(2) versus 1.34(3)).

The right panel of Fig.~\ref{fig:rr_pic_cylindrical} shows the $\rm\phi$-Z plane. Several artefacts appear as results of the masking of the central star, like the holes at $\rm\phi$ $\sim$ -60 and 120, but also the ones produced by the sliced nature of the datacube, which generates observable gaps in the data at angles of $\sim$ -150 and 30 degrees.
Even so, the structure of the equatorial ring is still appreciable, and in particular, it is possible to appreciate that the cloudy structures above and below the main ring are not continuous but fragmented. This is also evident in the 3D view of the shell (Fig.~\ref{fig:rr_pic_ppv_ppp}). The polar filaments in H$\rm\alpha$ span a wide range of $\rm\phi$ angles, which means they behave more like an extending ring than a bullet of material leaving the system. This does not seem to be the case for [O{\sc iii}] where the emission appears to be concentrated within a narrow range of $\rm\phi$ values.

\subsubsection{Fluxes and masses} \label{sec:flux_masses}

Together with their physical position, for each svoxel included in the final extracted 3D data, we also have their respective flux and associated uncertainty. This allows us to determine the total flux of the nova shell and its different components. The measured fluxes for H$\rm\alpha$, H$\rm\beta$ and [O{\sc iii}] are presented in Table~\ref{tab:flux_measurements}.
The fluxes were corrected for a reddening value of 0.034(17) mag, according to the 3D map of the interstellar medium provided by the Stilism website\footnote{\href{https://stilism.obspm.fr/}{Stilism website}} \citep{Stilism2017}. By using an extinction law parameter $R$ equal to 3.1 and the reddening law from \citet{Fitzpatrick99}, we determined correction factors of 1.08(4), 1.12(6), and 1.11(6) for the lines of H$\rm\alpha$, H$\rm\beta$, and [O{\sc iii}] $\rm\lambda$500.7 nm respectively.

\begin{table}[]
    \centering
    \caption{Fluxes and masses measured for the nova shell around RR Pic. The mass values given correspond to an electron temperature of 5\,000 K.}
    \label{tab:flux_measurements}
    \resizebox{\columnwidth}{!}{
    \begin{tabular}{l c c c}
    \hline \hline
    Shell & Flux                                 & Volume                & Mass        \\
          & [erg s$^{-1}$ cm$^{-2}$]             & [cm$^3$]              & [M$_\odot$] \\
    \hline
    H$\rm\alpha$    & & & \\
    Equatorial ring & 2.891(2)$\times$10$^{-14}$ & 3.04$\times$10$^{51}$ & 5.119(4)$\times$10$^{-5}$ \\
    SW filament     & 3.66(2)$\times$10$^{-16}$  & 5.32$\times$10$^{49}$ & 7.60(5)$\times$10$^{-7}$  \\
    NE filament     & 1.38(2)$\times$10$^{-16}$  & 2.17$\times$10$^{49}$ & 2.92(3)$\times$10$^{-7}$  \\
    Total           & 2.941(2)$\times$10$^{-14}$ & 3.12$\times$10$^{51}$ & 5.224(4)$\times$10$^{-5}$ \\
    H$\rm\beta$     & & & \\
    Equatorial ring & 4.42(5)$\times$10$^{-16}$  & 4.99$\times$10$^{49}$ & 1.43(1)$\times$10$^{-6}$  \\
    
    [O {\sc iii}]   & & & \\
    SW filament     & 1.47(1)$\times$10$^{-16}$  & 5.72$\times$10$^{49}$ & \\
    NE filament     & 8.97(9)$\times$10$^{-17}$  & 4.11$\times$10$^{49}$ & \\
    Total           & 2.37(1)$\times$10$^{-16}$  & 9.83$\times$10$^{49}$ & \\
    \hline
    \end{tabular}
    }
\end{table}

Together with the Balmer fluxes, we can provide rough estimates of the mass of the shell. The mass of a nova shell, $M_\mathrm{shell}$, can be estimated as:

\begin{equation}
    M_\mathrm{shell} = \mu \, m_\mathrm{p} \, n_\mathrm{p} \, V_\mathrm{shell} \, \epsilon ,\
    \label{eqn:mass_1}
\end{equation}

where $\mu$ corresponds to the mean molecular weight of the gas, $m_\mathrm{p}$ and $n_\mathrm{p}$ are the proton mass and density respectively, $V_\mathrm{shell}$ is the volume of the shell and $\epsilon$ is its filling factor. This approximation can be applied to planetary nebulae and nova shells \citep{Osterbrock2006}.

\begin{figure}
    \centering
    \includegraphics[width=1.0\columnwidth]{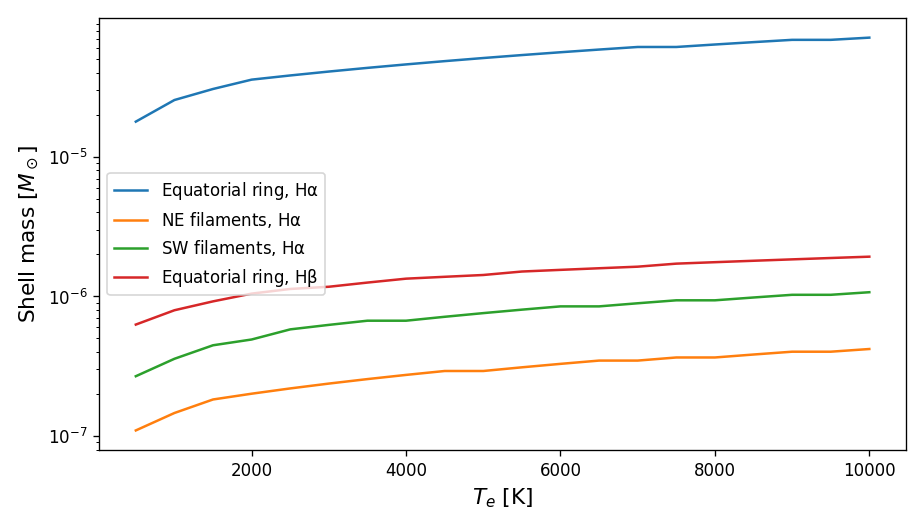}
    \caption{
    Shell masses as a function of electron temperature. The masses were determined by comparing the observed flux against theoretical values given a fixed electron temperature.
    }
    \label{fig:shellmass_1}
\end{figure}

Our data do not provide enough lines to constraint the electron density $n_\mathrm{e}$, electron temperatures $T_\mathrm{e}$ or $\epsilon$ within the gas, and therefore we need to search for an alternative method to determine the gas density.
As we are considering only the flux corresponding to the Balmer transitions, we can assume a gas composed purely of Hydrogen, which allows us to set $\mu$=1 and $n_\mathrm{p}$=$n_\mathrm{e}$. The electron density and temperature can then be estimated by comparing the observed energy emitted by the shell against its expected energy according to:

\begin{equation}
    4 \pi d^2 F_{\mathrm{obs}} = \epsilon(\mathrm{j},\mathrm{i}|n_\mathrm{e},T_\mathrm{e}) \, n_\mathrm{e} \, n_\mathrm{p} \, V_\mathrm{shell} ,\
    \label{eqn:mass_2}
\end{equation}

where $d$ corresponds to the distance to the system, $F_\mathrm{obs}$ to the observed flux and $\epsilon(\mathrm{j},\mathrm{i}|n_\mathrm{e},T_\mathrm{e})$ to the recombination emissivity for the Hydrogen transition j$\rightarrow$i given certain values for $n_\mathrm{e}$ and $T_\mathrm{e}$. For simplicity, we have assumed a filling factor $\epsilon$=1.
We define an array of values for $T_\mathrm{e}$ ranging from 500 to 10\,000 K in steps of 500 K. For each value of $T_\mathrm{e}$ we determine the value of $n_\mathrm{e}$ that best matches Eq.(~\ref{eqn:mass_2}).
We use the {\sc PyNeb Python} package \citep{Luridiana+2015_pyneb} to determine the values of $\epsilon(\mathrm{j},\mathrm{i}|n_\mathrm{e},T_\mathrm{e})$. These values correspond to the values published by \citet{Storey&Hummer1995}.
The volume of the shell can be directly determined from the 3D data extracted from the MUSE datacube as we can determine the volume that each svoxel occupies in the physical space.
Once the best value for $n_\mathrm{e}$ has been determined, we can derive a mass shell using Eq.(\ref{eqn:mass_1}). The shell masses computed as a function of $T_\mathrm{e}$ are presented in Fig.~\ref{fig:shellmass_1}, while the masses for an electron temperature of 5\,000 K, a representative value for the $T_\mathrm{e}$ of nova shells \citep[e.g.][]{Sahman18, Tappert23}, are presented in Table~\ref{tab:flux_measurements}.

\section{Discussion} \label{sec:discussion}

Historically the study of nova shells in the optical wavelengths has been limited to the use of narrow-band images and long-slit spectroscopy, either each technique on its own or a combination of the two. There are very few works in the literature that carried out IFS observations, these concerning the nova shells around V723 Cassiopeiae \citep{Lyke09}, HR Delphinus \citep{Moraes09}, the Helium nova V445 Puppis \citep{Macfarlane14}, V5668 Sagittarius \citep{Takeda22}, and QU Vulpecula \citep{Santamaria22b}. In these works the capabilities of the IFS observations have been used to study the spatio-kinematics, chemical composition, and morphology characteristics of these nova shells.

\subsection{Shell expansion}

In the present study of RR Pic, to develop a 3D view of the shell we have assumed it is expanding radially from the central star in all the observed lines. This assumption is supported by the comparison between the MUSE H$\rm\alpha$ image and the one published by GO98, in which several knots of material were identified in both images, being consistent with a radial, uniform expansion of the ejected material.
To our knowledge, there is currently no model for a non-radial expansion. One could in principle think that such non-radial expansion could be produced by strong magnetic fields and/or interaction with the secondary star, but to date, no evidence for such an expanding shell has been reported. On the other hand, the assumption of radial expansion is supported by multi-epoch observations of several nova shells \citep[e.g.][]{Liimets12, Harvey16, Santamaria20}.

The assumption of a uniform, radially expanding shell was used by \citet{Takeda22} to create a 3D view of the young shell around the system V5668 Sgr. When compared with our approach, the main difference is that we have left open the possibility of including different models for the expansion velocity history of the material, instead of the free expansion only. This, in principle, is useful for older shells where deceleration of the expanding material is expected to occur. Our test case, RR Pic, is one of the oldest novae recorded, so it was expected from our side to find some signals of deceleration within the shell. 

The overall geometry of the RR Pic nova shell previously reported as consisting of an equatorial ring around the system and polar filaments leaving the system in an orthogonal direction to the ring is preserved, as can be seen in the MUSE H$\rm\alpha$ image (Fig.~\ref{fig:comparison_go98_muse}). From the measurements of the radial distance of the several knots identified in the image, in both MUSE and GO98 images, we found that the shell is consistent with a still free-expanding phase (Fig.~\ref{fig:rr_pic_shell_expansion}).
It has been proposed that nova shells start to show signals of deceleration decades after the nova eruption, decreasing its initial expansion velocity by half $\sim$ 75 years after \citep{Duerbeck87a}. However, recent studies have contradicted this idea \citep{Liimets12, Santamaria20}. In particular, the recent studies on the nova shells around T Aur ($\sim$125 years) and V476 Cyg ($\sim$100 years) reveal these shells are still expanding free \citep{Santamaria20}, which together with the results on RR Pic ($\sim$96 years), indicates that nova shells can expand freely for more than a century after the nova eruption.

\subsection{Interaction with the ISM} \label{sec:discussion_interaction_ism}

In the study of supernova remnants, it is expected that the deceleration occurs when the material swept up by the expanding shell has a similar mass to the initially ejected one, the moment in which the expanding shell will enter a new phase called the Sedov-Taylor phase \citep{Reynolds08}. By applying the same principle to nova shells, it could be possible to estimate the time required for the shell to reach this phase.
From the mass estimated in Section~\ref{sec:flux_masses} we conclude most of the mass of the shell is confined to the equatorial ring.
Using the mass estimated at $T_\mathrm{e}$=5\,000 K corresponding to $\sim$5$\times$10$^{-5}$ M$_\odot$ (Sect.~\ref{sec:flux_masses}), and the opening angle of the ring of $\sim$30 deg (Sect.~\ref{sec:3d_geometrical_measurements}), we can compute the mass drag by the equatorial ring, assuming an ISM with homogeneous density.
As RR Pic is relatively close to the Sun ($\sim$500 pc), for the density value of the Interstellar medium we use a value representative of the local interstellar medium, $n_\mathrm{H}$=0.1 cm$^{-3}$ \citep{Frisch+11}. 
Assuming a spherical geometry corrected by the opening angle of the expanding equatorial ring, and a uniform expansion velocity of 370 km s$^{-1}$ the time required to drag a mass equivalent to the measured mass of the ring will be approximately $\sim$650 yrs.
If we increase the ISM density by one order of magnitude, the time required would decrease up to a value of $\sim$300 years. If instead we decrease the ISM density by a factor of ten, then this time would increase up to $\sim$1400 years.
These numbers suggest that the equatorial ring will last centuries expanding freely before reaching the Sedov-Taylor phase and starting to show signals of deceleration.
This potentially has significant consequences for the determination of ages of the shell of unrecorded nova eruptions \citep[see for example][]{Shara+17_ATCnc, Santamaria19}

The situation could be very different for the case of the polar filaments, as they present much lower masses than the equatorial ring. While these lower masses of the knots can be compensated with a smaller dragged volume, preventing an earlier deceleration compared to the equatorial ring, other possible interactions could occur.
As the knot advances through a less dense medium, some processes like Rayleigh-Taylor instabilities could start to play a role in the evolution of the knot. It is unlikely that these instabilities will be able to change the direction of the advancing knot, but they could affect the knot's surface and its surface brightness. This could be the cause behind the apparent acceleration observed in the NE filaments knots between the GO98 and MUSE images (Sect.~\ref{sec:comparison_go98}).

\begin{figure}
    \centering
    \includegraphics[width=1.0\columnwidth]{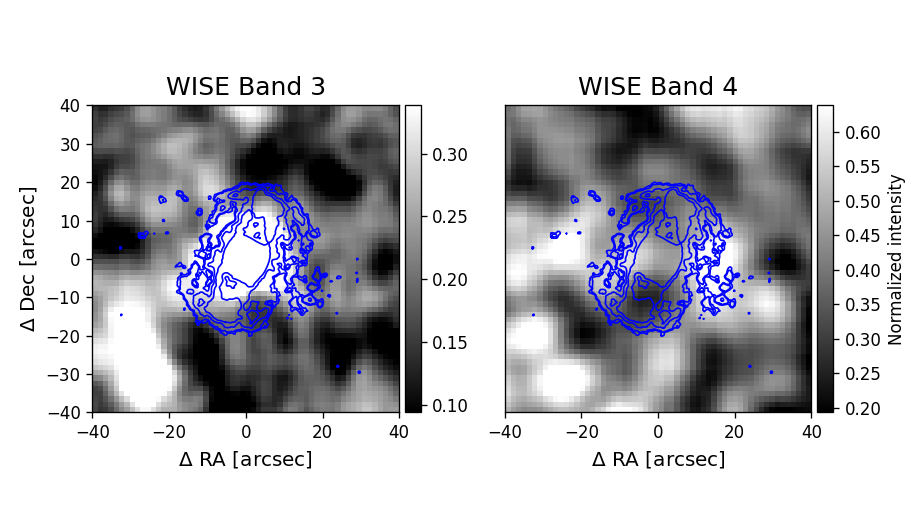}
    \caption{Normalized intensity maps of WISE bands 3 and 4 around the position of RR Pic. The nova shell observed with MUSE is presented with blue contours.}
    \label{fig:wise}
\end{figure}

To test this idea, we checked the available data of the Wide-Field Infrared Explored Survey, WISE \citep{WISE2010}. The WISE filter at 12$\mathrm{\text\textmu m}$ (WISE band 3) is sensitive to the Poly-Aromatic Hydrocarbures, ubiquitous tracer of the ISM material, while the WISE filter at 22$\mathrm{\text\textmu m}$ (WISE band 4) is sensitive to small and large dust grain particles.
The normalized intensity map of the WISE bands 3 and 4 for a region of 80 arcsec size centred at the position of RR Pic is presented in Fig.~\ref{fig:wise}. In the figure, the nova shell as it was observed by MUSE in H$\rm\alpha$ is presented in blue contours. The data show that in the WISE Band 3, there is a zone of higher intensity in the NE direction of the shell with respect to the intensities observed in the SW direction. This could indicate that the NE filament of the shell is entering a zone where the ISM has a relatively higher density. The WISE Band 4 on the other hand presents a difference in the intensity observed in the NE-SW (polar filament) axis with respect to the SE-NW (equatorial ring) direction. The intensity in the axis of the polar filament appears to be higher than the intensity across the equatorial ring, suggesting a higher amount of dust in the ISM in the direction of the expanding filaments.
We must note however that in spite of the seductive coincidence observed between the equatorial ring and polar filaments observed in the WISE band 4 image, we do not have a way to estimate the distance to the features in the WISE images, and therefore, we cannot confirm whether the aforementioned possibilities are real or just the result of a projection effect.

Additional evidence for interaction between the nova shell and the ISM could come in the form of shock waves, which should increase the luminosity of the shell in X-rays and radio wavelengths \citep[][and references therein]{Chomiuk21_novareview}.
Chandra observations of the RR Pic's nova shell showed an extended X-ray emission source which follows the equatorial ring, indicating a shock between the ring and the surrounding ISM \citep{Balman2006}. The cloudy structures observed in the equatorial ring (Fig.~\ref{fig:rr_pic_cylindrical}) could find their origin as the result of the aforementioned interaction.
No X-ray detection regarding the polar filaments has been reported.

\subsection{Nova shell geometry}

Our results indicate that the equatorial and polar materials are expanding at different velocities, with the polar filaments expanding faster. This is consistent with the idea of two-phase ejections proposed by \citet{Aydi20}, where the material in the poles of the CV leaves the system faster, as there is less material in this direction to oppose the ejected material.
The apparent chemical composition of these two regions is different also, with the equatorial ring traced by Hydrogen mainly, and the polar filaments by Hydrogen and [O{\sc iii}] (Fig.~\ref{fig:rr_pic_image_velocity}). Weak traces of [N{\sc ii}] can also be observed in the NE polar filaments. These differences between the ring and filaments could be due to differences in abundances, but also to differences in the densities and temperatures of these environments. This latter scenario would imply a denser equatorial ring, which is in agreement with results provided by hydrodynamical simulations \citep{Booth+2016_RSOph}.
What was previously unnoticed is that the NE filaments are apparently expanding faster than the SW filament by a non-negligible factor ($\sim$ 40\% faster). As we could not find strong evidence for interaction between these filaments and the surrounding ISM, we must consider the possibility that the observed differences in expansion velocity must come from the nova explosion itself. The observed asymmetry in the expansion velocity could be explained then by a difference in the ejected mass, a difference between the energy of the nova eruption on the poles of the WD, or a combination of both.

From our mass estimation, we can observe that there is a difference ratio of $\sim$2.5 between the SW and NE filament masses. This ratio is similar to the differences in the observed fluxes (Table~\ref{tab:flux_measurements}). It is worth noting that the NE filament corresponds to the receding filament, and therefore its lower flux with respect to the SW filaments could be explained by self-absorption of the shell caused by the formation of grain dust \citep{Shore+2018_dust}. This idea has been proposed as a solution to explain the observed asymmetry in flux between blue and red peaks in the spectra of young, dusty classical novae. While a certain amount of self-absorption is possible within the system, it is very unlikely that this could explain the observed difference in flux between the NE and SW filament, as the shell has expanded considerably, departing from the scenario proposed in \citet{Shore+2018_dust}. Furthermore, the light curve of RR Pic did not show evidence for dust formation \citep{Strope10}.
Therefore, we can conclude that the differences in flux must be intrinsic to the filaments. We can compare the kinetic energy of both filaments by using the expansion velocities determined at the end of Section~\ref{sec:expansion_history}. This leads to kinetic energies of the order of 1.7$\times$10$^{42}$ erg for the SW and 1.0$\times$10$^{42}$ erg for the NE filament, giving a ratio between the SW and NE of 1.7. The observed differences in fluxes, masses and kinetic energies suggest an intrinsic origin for the observed asymmetry, the genesis for which could be found in the nova eruption itself.
While the origin of such asymmetries is unclear, it has been hypothesized that they may be an important factor in the long-term evolution of CVs \citep{Nelemans+16, Schaefer+19_QZAur}.

From our 3D model, we can observe in Fig.~\ref{fig:rr_pic_cylindrical} the allowed and forbidden transitions trace different regions of the shell, with Hydrogen tracing the equator and the poles regions of the system while Oxygen follows the tropical latitudes where no Hydrogen is observed. We noticed that the distribution of both emitters is best fitted in the $\rm\rho$-Z plane by different ellipses, with the Hydrogen being best fitted by a prolate ellipse with an axial ratio of 1.35(2), and the Oxygen by a prolate ellipse with an axial ratio of 0.92(3).
The morphology of nova shells correlates with their expansion speed, being more spherical for faster shells \citep{Santamaria22a}. RR Pic was considered a slow nova, with expansion velocities of the order of 400 km s$^{-1}$ and a $t_3$ of 122 d, the reason why we would expect the shell to have a prolate shape. And in fact, that is what we observe in H$\rm\alpha$. However, the axial ratio differs when we use the data from the projected image (Fig.~\ref{fig:rr_pic_image_velocity}) or those from the 3D reconstruction (Fig.~\ref{fig:rr_pic_cylindrical}). In the first case we can determine the axial ratio from the polar and equatorial radial extension (Fig.~\ref{fig:rr_pic_shell_expansion}), which leads to an axial ratio of 1.445(2), while when we measured through the fit of the 3D data, we obtain a value of 1.35(3). Differences in the geometry derived from image and IFS have been reported before \citep{Santamaria22b}, which highlights the importance and advantage of using IFS data.
In the case of [O{\sc iii}], the best fit suggests an oblate shape with an axial ratio smaller than one. Because the Oxygen material is confined to the tropical latitudes of the shell only, the fit does not have any constraints in the equatorial region. As a consequence, it favours an oblate shape due to the aperture angle observed in the [O{\sc iii}] emission.

\subsection{CV properties from the nova shell}

The 3D data also allow us to find the circular ring that best fits the geometry of the equatorial ring around the system, from which we could determine its inclination to be 73.6$\pm$0.3 deg. Assuming a co-planar geometry, we can extrapolate this result to the inclination of the system itself.
The light curve of RR Pic shows an orbital hump of variable amplitude, and possibly also shallow eclipses, with the latter being dependent on the structure of the accretion disc \citep{Schmidtobreick08, Fuentes-Morales18}. This suggests that the inclination of the system must be at an intermediate angle that allows for shallow eclipses only at certain states of the accretion disc.
\citet{Sion17} used Far Ultraviolet Spectroscopic Explorer (FUSE) data to estimate an inclination of 60 degrees for the accretion disc and a mass of 1 M$_\odot$ for the WD. These physical parameters were obtained by fitting the observed continuum of the disc and, as pointed out by the authors, would be different when fitting instead the absorption lines in the spectrum. This indicates that there is still room for improvement of the disc model implemented by the authors, which could be the reason behind the discrepancy between their and our value for the system inclination.

Another possibility is that the inclination of the shell and the inclination of the disc are actually different. However, a coplanar geometry between the orbital plane and the equatorial outflow is supported by hydrodynamical simulations of the ejected material around the symbiotic system RS Oph \citep{Booth+2016_RSOph}. If this were the case also in RR Pic, then a tilted disc would be required to explain the differences in inclination regarding the equatorial ring and the accretion disc.
A tilted disc has been indicated to be the cause for negative superhumps, which correspond to periodic modulations of the CV light curves with a period shorter than the orbital, and have been observed in classical novae and nova-likes system \citep[See for example][]{Fuentes-Morales18, Ilkiewicz21_AQMen}. However, no negative superhumps have been reported on RR Pic \citep{Fuentes-Morales18}, which argues against the possibility of a tilted disc within the system.
For the mentioned reasons, we concluded that the inclination of the system should be closer to our result of $\sim$70 degrees.

The 3D view of the data shows the equatorial ring in RR Pic nova shell to be composed of a continuous ring where most of the emission comes from, but also of several cloudy structures above and below the main ring, which extends to a height of $\pm$4\,000 au, in a configuration that is reminiscent of a bow shock (Fig.~\ref{fig:rr_pic_cylindrical}, left panel). The origin of these clouds could be intrinsic to the nova eruption, or the result of the interaction between the expanding material and the static ISM. In the first scenario, it is expected that the same structure is present in the image published by GO98. Unfortunately, the quality of the published image makes it difficult to appreciate the details of these clouds, but it appears that some structures in the northern part of the shell have persisted throughout the years, with no significant change to their appearance, supporting an intrinsic formation scenario. According to the simulations of \citet{Porter98}, tropical rings could be formed in the shell, whose height with respect to the orbital plane depends on the WD rotation. Such tropical rings could present a potential place of origin for the observed clouds.

From \citet{Porter98}, the rotational velocity at the surface of the WD $v_{\mathrm{\phi}}$ is given by: 

\begin{equation}
    v_\mathrm{\phi} = f(R_\mathrm{WD}, \pi/2) \sqrt{\frac{G M_\mathrm{WD}}{R_\mathrm{WD}}}
    \label{eqn:v_phi_from_Porter98}
\end{equation}

where $M_\mathrm{WD}$ and $R_\mathrm{WD}$ are the mass and radius of the WD respectively. The dimensionless factor $f(R_\mathrm{WD},\pi/2)$ corresponds to the ratio between the WD rotation period and the Keplerian rotation period of a particle at the radius of the WD and can be related to the height of the tropical rings. While \citet{Porter98} did not provide values of $f(R_\mathrm{WD},\pi/2)$ as a function of the tropical rings opening angle $\rm\theta$, some information can be extracted from their Figure 1. The extracted angles $\rm\theta$ and their respective value for $f(R_\mathrm{WD},\pi/2)$ are presented in Table~\ref{tab:porter_angles}.

\begin{table}[]
    \centering
    \caption{Tropical rings opening angle $\rm\theta$ as a function of the factor $f(R_\mathrm{WD},\pi/2)$. Extracted from Figure 1 of \citet{Porter98}.}
    \label{tab:porter_angles}
    \begin{tabular}{cc}
        \hline
        \hline
        $\rm\theta$ [deg] & $f(R_\mathrm{WD},\pi/2)$ \\
        \hline
        25 & 0.0 \\
        37 & 0.5 \\
        77 & 0.7 \\
        \hline
    \end{tabular}
\end{table}

Using the opening angle of 30 deg for the cloudy structures (Section~\ref{sec:3d_geometrical_measurements}) and by linear interpolation of the data provided in Table~\ref{tab:porter_angles}, we obtain a value for $f(R_\mathrm{WD},\pi/2)$ of 0.21.

With a WD mass of 1 M$_\odot$ \citep{Sion17} and a WD radius of 0.008 R$_\odot$ that results from the theoretical mass-radius relation for WDs from \citet{Fontaine01}, $v_\mathrm{\phi}$ then results in 1025 km s$^{-1}$, which translates into a spin period for the WD of 31 seconds. In most non-magnetic CVs, the reported value for the WD $v \sin (i)$ is below 400 km s$^{-1}$ \citep{Sion99}, while for RR Pic, our estimate would result in $\sim$980 km s$^{-1}$. This would place RR Pic in a row with intermediate polar systems like WZ Sge and CTCV J2056-3014 which possess spin periods of 28s and 29s respectively \citep{Lasota99, LopezdeOliveira20_CTCVJ2056-3014}.

\subsection{Possible evolution of the nova shell}

\begin{figure}
    \centering
    \includegraphics[width=1.0\columnwidth]{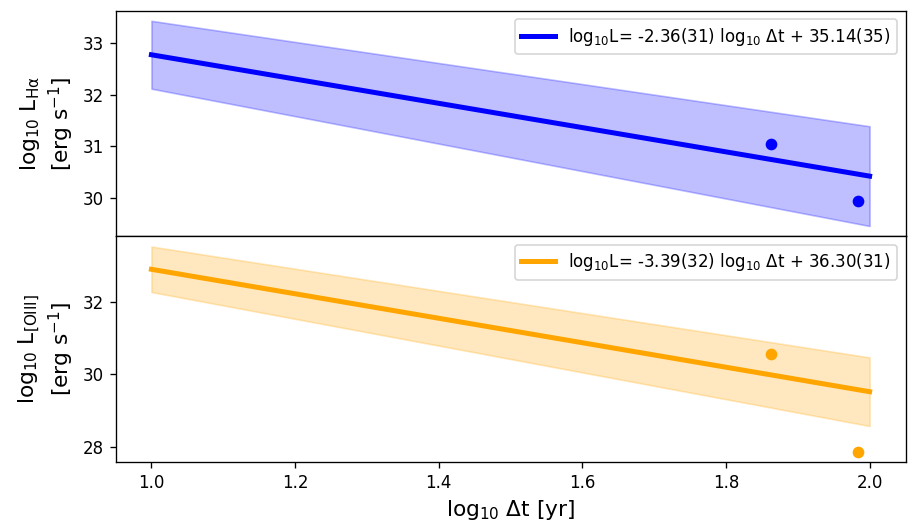}
    \caption{
    Expected H$\rm\alpha$ (blue) and [O{\sc iii}] (orange) luminosity evolution for RR Pic according to \citet{Tappert20}. The dots represent the measured flux for RR Pic by \citet{Downes01} and this work. The H$\rm\alpha$ luminosity is consistent with the expected value given the uncertainties, while the [O{\sc iii}] luminosity appears to be fainter than what is expected.
    } 
    \label{fig:shell_luminosity}
\end{figure}

Understanding how the nova shells evolve is key to fully comprehending the origin of the ancient nova shells and their implications for the CV evolution.
The luminosity evolution of nova shells in H$\rm\alpha$ and [O{\sc iii}] as a function of the time since nova eruption was studied by \citet{Downes01} and \citet{Tappert20}, with the latter using the light curve classification used in the work of \citet{Strope10}. RR Pic was classified as a member of the jitter subgroup. The expected evolution for this group in the log$_{10}$\,L - log$_{10}$\,$\Delta t$ diagram is presented in Fig~\ref{fig:shell_luminosity}, together with the luminosity reported in \citet{Downes01} ($\Delta t \sim$ 74 years) and from the MUSE data ($\Delta t \sim$ 96 years).

The luminosity from MUSE in H$\rm\alpha$ is consistent with the expected decline within the given uncertainties, but the [O{\sc iii}] luminosity appears to be fainter with respect to the expected value.
From their Figure 3, it is clear that the fit proposed by \citet{Tappert20} is dominated by the nova with the largest number of points (HR Del), while at least one other system in this group, PW Vul, appears to follow a steeper decline.
It is therefore feasible that RR Pic also follows a steeper decline in [O{\sc iii}] than the one indicated by the fit.
Furthermore, \citet{Tappert20} suggest that at the late stage of their evolution, nova shells reach the 'ancient nova plateau', a region where the luminosity of ancient shells appears to maintain a constant value regarding the time since their eruption. It is not clear however whether or not all shells reach this stage and if they do, at what time after the nova eruption. Future observations of RR Pic's nova shell during the next decades will help to answer these questions.

Focusing on the H$\rm\alpha$ flux, most of it is concentrated on the equatorial ring and just a minor fraction in the polar filaments. It is very likely then that in the future the filaments will disperse sooner than the ring, ending with a very similar geometry to the ancient shells of AT Cnc \citep{Shara+17_ATCnc} and IPHASX J210204.7+471015 \citep{Santamaria19}. These ancient shells present incomplete ring structures that have fragmented into several blobs of material, with both of them being dominated by forbidden transitions of [O{\sc iii}] and [N{\sc ii}].
It is certainly possible that the equatorial ring currently observed in RR Pic evolves into a geometry similar to the ancient shells mentioned above. As the ring expands and interacts with the ISM it should start to fragment while its density decreases causing the ring to start transitioning towards being dominated by forbidden transitions.
The ages of AT Cnc and IPHASX J210204.7+471015 are estimated to be at least 200 years old and $\sim$170 years old respectively. Assuming the proposed evolution scenario for the equatorial ring of RR Pic is correct, then such a transition should happen within the next century.

\section{Conclusions} \label{sec:conclusions}

In this work, we have shown for the first time the study and analysis of a nova shell observed using the MUSE instrument from Paranal, ESO. We have presented our methodology for extracting the 3D data of a nova shell from the MUSE datacube, and how we transformed the observed PPV space into a physical PPP space.

As a test case for our methodology, we have presented the observations regarding the classical nova RR Pictoris (Nova Pic 1925). 
The nova shell around the binary system was clearly observed in the H$\rm\alpha$ and H$\rm\beta$ lines, as well as in the [O{\sc iii}] $\rm\lambda$492.1 and $\rm\lambda$500.7 nm lines and very faintly in the [N{\sc ii}] $\rm\lambda$684.8 nm line.
The overall structure observed by GO98 in their H$\rm\alpha$ narrow-band image, consisting of an equatorial ring and polar filaments, is still preserved. From a comparison with those previous observations, we have concluded that the shell is likely to continue in its free-expansion phase.
The expansion rates differ from the equatorial ring and the polar filaments, with the former expanding at a rate of 0.154(1) arcsec yr$^{-1}$. The polar filaments are expanding at different rates with the NE filaments expanding at 0.26(1) arcsec yr$^-1$ and the SW filaments expanding at 0.199(5) arcsec yr$^-1$.

The observed spectrum shows that the equatorial ring is dominated by Balmer emission, while the polar filaments are observed mainly in the [O{\sc iii}] lines and to a lesser extent also in Balmer lines. The total H$\rm\alpha$ flux measured in the polar filaments corresponds to $\sim$1\% of the equatorial ring flux.
Comparing the polar filaments, we noted that their fluxes are not equal, with the SW filaments being $\sim$2.5 and $\sim$1.7 times brighter than the NE filament for the lines of H$\rm\alpha$ and [O{\sc iii}] $\rm\lambda$500.7 nm, respectively.
From the total measured H$\rm\alpha$ flux, we estimate the mass of the shell to be $\sim$5$\times$10$^{-5}$ M$_\odot$.

The extracted 3D view of the data allowed us to better constrain the geometry of the nova shell, in particular concerning its equatorial ring. We have fitted the ring assuming a circular geometry to determine its radial extension, inclination, and position angle. Our results indicate a ring with a radius of $\sim$8\,000 au tracing the inner and denser parts of it, but extending between $\sim$6\,000 and $\sim$10\,000 au when considering the fainter regions of the ring. The inclination of the ring and its position angle are well constrained to $\sim$74 and $\sim$155 deg, respectively. We argue that the inclination value for the ring corresponds to the inclination of the binary.

The mean radial profile of the equatorial ring shows a shape that is reminiscent of a bow shock, with clouds of material extending up and below the equatorial ring, with a height of $\sim$4\,000 AU. The polar filaments extend asymmetrically with the northern one extending further at a maximum distance of $\sim$15\,000 AU. If we ignore the northern filament, the overall 3D structure of the shell observed in H$\rm\alpha$ is well fitted by an ellipsoid with an axial ratio of 1.34(3), with the semi-major axis following the polar axis.
The H$\rm\alpha$ and [O{\sc iii}] emissions trace different regions within the shell, with the latter tracing the gap between the equatorial ring and the polar material. This indicates a difference in density across the shell related to its polar angle.

Exploring the possibility that the nova shell is experiencing an interaction with the surrounding ISM, we have analysed the WISE public data in the search for evidence supporting this idea, but we find the results to be inconclusive.
Theorizing the origin of the observed cloudy material within the equatorial ring, we have explored the possibility that they correspond to the tropical rings predicted by the hydrodynamical simulations of \citep{Porter98}. Based on that assumption, we have estimated the spin period of the WD to be $\sim$31 seconds.

Lastly, we have speculated about the future of the nova shell around RR Pic. We compared the current observed luminosity with the expected values according to the luminosity evolution proposed by \citep{Tappert20}. The result is consistent for the H$\rm\alpha$ luminosity, but the [O{\sc iii}] decline is steeper compared with the expected one.
As the RR Pic's nova shell continues to evolve, we conjecture that the filaments will disperse into the ISM, leaving only the equatorial ring in a geometry that is similar to that currently observed in some ancient nova shells.

This work has shown the advantages of using IFS in the study of nova shells, while also pointing to the interesting features observed in the nova shell around RR Pic. Many of the results and ideas presented in this work can be tested with further and repeated observations of the shell.

\begin{acknowledgements}

The raw and reduced MUSE datacubes used in this work can be freely accessed from the ESO archive.
The CSV files containing the extracted data of the shell can be accessed via Github\footnote{\url{https://github.com/LCeledon/Novashell3D_MUSE}}.

LC acknowledges economic support from ANID-Subdireccion de capital humano/doctorado nacional/2022-21220607 and ESO-Vitacura.
The authors want to thank Fuyan Bian for organizing the Apocalypse Filler Program on UT4 which has been used to apply for these data.

\end{acknowledgements}

\bibliographystyle{aa} 
\bibliography{file.bib} 

\begin{thebibliography}{75}
\expandafter\ifx\csname natexlab\endcsname\relax\def\natexlab#1{#1}\fi

\bibitem[{{Anupama} \& {Kantharia}(2005)}]{Anupama&Kantharia05}
{Anupama}, G.~C. \& {Kantharia}, N.~G. 2005, \aap, 435, 167

\bibitem[{{Aydi} {et~al.}(2020){Aydi}, {Chomiuk}, {Izzo}, {Harvey},
  {Leahy-McGregor}, {Strader}, {Buckley}, {Sokolovsky}, {Kawash}, {Kochanek},
  {Linford}, {Metzger}, {Mukai}, {Orio}, {Shappee}, {Shishkovsky}, {Steinberg},
  {Swihart}, {Sokoloski}, {Walter}, \& {Woudt}}]{Aydi20}
{Aydi}, E., {Chomiuk}, L., {Izzo}, L., {et~al.} 2020, \apj, 905, 62

\bibitem[{{Bacon} {et~al.}(2010){Bacon}, {Accardo}, {Adjali}, {Anwand},
  {Bauer}, {Biswas}, {Blaizot}, {Boudon}, {Brau-Nogue}, {Brinchmann},
  {Caillier}, {Capoani}, {Carollo}, {Contini}, {Couderc}, {Daguis{\'e}},
  {Deiries}, {Delabre}, {Dreizler}, {Dubois}, {Dupieux}, {Dupuy}, {Emsellem},
  {Fechner}, {Fleischmann}, {Fran{\c{c}}ois}, {Gallou}, {Gharsa}, {Glindemann},
  {Gojak}, {Guiderdoni}, {Hansali}, {Hahn}, {Jarno}, {Kelz}, {Koehler},
  {Kosmalski}, {Laurent}, {Le Floch}, {Lilly}, {Lizon}, {Loupias}, {Manescau},
  {Monstein}, {Nicklas}, {Olaya}, {Pares}, {Pasquini}, {P{\'e}contal-Rousset},
  {Pell{\'o}}, {Petit}, {Popow}, {Reiss}, {Remillieux}, {Renault}, {Roth},
  {Rupprecht}, {Serre}, {Schaye}, {Soucail}, {Steinmetz}, {Streicher}, {Stuik},
  {Valentin}, {Vernet}, {Weilbacher}, {Wisotzki}, \& {Yerle}}]{Bacon10}
{Bacon}, R., {Accardo}, M., {Adjali}, L., {et~al.} 2010, in Society of
  Photo-Optical Instrumentation Engineers (SPIE) Conference Series, Vol. 7735,
  Ground-based and Airborne Instrumentation for Astronomy III, ed. I.~S.
  {McLean}, S.~K. {Ramsay}, \& H.~{Takami}, 773508

\bibitem[{{Bacon} {et~al.}(2016){Bacon}, {Piqueras}, {Conseil}, {Richard}, \&
  {Shepherd}}]{Bacon16}
{Bacon}, R., {Piqueras}, L., {Conseil}, S., {Richard}, J., \& {Shepherd}, M.
  2016, {MPDAF: MUSE Python Data Analysis Framework}, Astrophysics Source Code
  Library, record ascl:1611.003

\bibitem[{{Balman}(2005)}]{Balman05}
{Balman}, {\c{S}}. 2005, \apj, 627, 933

\bibitem[{{Balman}(2006)}]{Balman2006}
{Balman}, {\c{S}}. 2006, Advances in Space Research, 38, 2840

\bibitem[{{Bertin} \& {Arnouts}(1996)}]{Sextractor}
{Bertin}, E. \& {Arnouts}, S. 1996, \aaps, 117, 393

\bibitem[{{Booth} {et~al.}(2016){Booth}, {Mohamed}, \&
  {Podsiadlowski}}]{Booth+2016_RSOph}
{Booth}, R.~A., {Mohamed}, S., \& {Podsiadlowski}, P. 2016, \mnras, 457, 822

\bibitem[{{Capitanio} {et~al.}(2017){Capitanio}, {Lallement}, {Vergely},
  {Elyajouri}, \& {Monreal-Ibero}}]{Stilism2017}
{Capitanio}, L., {Lallement}, R., {Vergely}, J.~L., {Elyajouri}, M., \&
  {Monreal-Ibero}, A. 2017, \aap, 606, A65

\bibitem[{{Castro Segura} {et~al.}(2021){Castro Segura}, {Knigge},
  {Acosta-Pulido}, {Altamirano}, {del Palacio}, {Hernandez Santisteban},
  {Pahari}, {Rodriguez-Gil}, {Belardi}, {Buckley}, {Burleigh}, {Childress},
  {Fender}, {Hewitt}, {James}, {Kuhn}, {Kuin}, {Pepper}, {Ponomareva},
  {Pretorius}, {Rodr{\'\i}guez}, {Stassun}, {Williams}, \&
  {Woudt}}]{CastroSegura21}
{Castro Segura}, N., {Knigge}, C., {Acosta-Pulido}, J.~A., {et~al.} 2021,
  \mnras, 501, 1951

\bibitem[{{Chomiuk} {et~al.}(2021){Chomiuk}, {Metzger}, \&
  {Shen}}]{Chomiuk21_novareview}
{Chomiuk}, L., {Metzger}, B.~D., \& {Shen}, K.~J. 2021, \araa, 59, 391

\bibitem[{{Downes} \& {Duerbeck}(2000)}]{DownesDuerbeck00}
{Downes}, R.~A. \& {Duerbeck}, H.~W. 2000, \aj, 120, 2007

\bibitem[{{Downes} {et~al.}(2001){Downes}, {Duerbeck}, \&
  {Delahodde}}]{Downes01}
{Downes}, R.~A., {Duerbeck}, H.~W., \& {Delahodde}, C.~E. 2001, Journal of
  Astronomical Data, 7, 6

\bibitem[{{Duerbeck}(1987{\natexlab{a}})}]{Duerbeck87CatalogueNovae}
{Duerbeck}, H.~W. 1987{\natexlab{a}}, \ssr, 45, 1

\bibitem[{{Duerbeck}(1987{\natexlab{b}})}]{Duerbeck87a}
{Duerbeck}, H.~W. 1987{\natexlab{b}}, \apss, 131, 461

\bibitem[{{Duerbeck}(1987{\natexlab{c}})}]{Duerbeck87d}
{Duerbeck}, H.~W. 1987{\natexlab{c}}, The Messenger, 50, 8

\bibitem[{{Evans} {et~al.}(1992){Evans}, {Bode}, {Duerbeck}, \&
  {Seitter}}]{Evans92_RRPic}
{Evans}, A., {Bode}, M.~F., {Duerbeck}, H.~W., \& {Seitter}, W.~C. 1992,
  \mnras, 258, 7P

\bibitem[{{Fang} {et~al.}(2014){Fang}, {Guerrero}, {Marquez-Lugo}, {Toal{\'a}},
  {Arthur}, {Chu}, {Blair}, {Gruendl}, {Hamann}, {Oskinova}, \&
  {Todt}}]{Fang+2014}
{Fang}, X., {Guerrero}, M.~A., {Marquez-Lugo}, R.~A., {et~al.} 2014, \apj, 797,
  100

\bibitem[{{Fitzpatrick}(1999)}]{Fitzpatrick99}
{Fitzpatrick}, E.~L. 1999, \pasp, 111, 63

\bibitem[{{Fontaine} {et~al.}(2001){Fontaine}, {Brassard}, \&
  {Bergeron}}]{Fontaine01}
{Fontaine}, G., {Brassard}, P., \& {Bergeron}, P. 2001, \pasp, 113, 409

\bibitem[{{Frisch} {et~al.}(2011){Frisch}, {Redfield}, \& {Slavin}}]{Frisch+11}
{Frisch}, P.~C., {Redfield}, S., \& {Slavin}, J.~D. 2011, \araa, 49, 237

\bibitem[{{Fuentes-Morales} {et~al.}(2018){Fuentes-Morales}, {Vogt}, {Tappert},
  {Schmidtobreick}, {Hambsch}, \& {Vu{\v{c}}kovi{\'c}}}]{Fuentes-Morales18}
{Fuentes-Morales}, I., {Vogt}, N., {Tappert}, C., {et~al.} 2018, \mnras, 474,
  2493

\bibitem[{{Gaia Collaboration} {et~al.}(2021){Gaia Collaboration}, {Brown},
  {Vallenari}, {Prusti}, {de Bruijne}, {Babusiaux}, {Biermann}, {Creevey},
  {Evans}, {Eyer}, {Hutton}, {Jansen}, {Jordi}, {Klioner}, {Lammers},
  {Lindegren}, {Luri}, {Mignard}, {Panem}, {Pourbaix}, {Randich}, {Sartoretti},
  {Soubiran}, {Walton}, {Arenou}, {Bailer-Jones}, {Bastian}, {Cropper},
  {Drimmel}, {Katz}, {Lattanzi}, {van Leeuwen}, {Bakker}, {Cacciari},
  {Casta{\~n}eda}, {De Angeli}, {Ducourant}, {Fabricius}, {Fouesneau},
  {Fr{\'e}mat}, {Guerra}, {Guerrier}, {Guiraud}, {Jean-Antoine Piccolo},
  {Masana}, {Messineo}, {Mowlavi}, {Nicolas}, {Nienartowicz}, {Pailler},
  {Panuzzo}, {Riclet}, {Roux}, {Seabroke}, {Sordo}, {Tanga}, {Th{\'e}venin},
  {Gracia-Abril}, {Portell}, {Teyssier}, {Altmann}, {Andrae}, {Bellas-Velidis},
  {Benson}, {Berthier}, {Blomme}, {Brugaletta}, {Burgess}, {Busso}, {Carry},
  {Cellino}, {Cheek}, {Clementini}, {Damerdji}, {Davidson}, {Delchambre},
  {Dell'Oro}, {Fern{\'a}ndez-Hern{\'a}ndez}, {Galluccio}, {Garc{\'\i}a-Lario},
  {Garcia-Reinaldos}, {Gonz{\'a}lez-N{\'u}{\~n}ez}, {Gosset}, {Haigron},
  {Halbwachs}, {Hambly}, {Harrison}, {Hatzidimitriou}, {Heiter},
  {Hern{\'a}ndez}, {Hestroffer}, {Hodgkin}, {Holl}, {Jan{\ss}en}, {Jevardat de
  Fombelle}, {Jordan}, {Krone-Martins}, {Lanzafame}, {L{\"o}ffler}, {Lorca},
  {Manteiga}, {Marchal}, {Marrese}, {Moitinho}, {Mora}, {Muinonen}, {Osborne},
  {Pancino}, {Pauwels}, {Petit}, {Recio-Blanco}, {Richards}, {Riello},
  {Rimoldini}, {Robin}, {Roegiers}, {Rybizki}, {Sarro}, {Siopis}, {Smith},
  {Sozzetti}, {Ulla}, {Utrilla}, {van Leeuwen}, {van Reeven}, {Abbas}, {Abreu
  Aramburu}, {Accart}, {Aerts}, {Aguado}, {Ajaj}, {Altavilla}, {{\'A}lvarez},
  {{\'A}lvarez Cid-Fuentes}, {Alves}, {Anderson}, {Anglada Varela}, {Antoja},
  {Audard}, {Baines}, {Baker}, {Balaguer-N{\'u}{\~n}ez}, {Balbinot}, {Balog},
  {Barache}, {Barbato}, {Barros}, {Barstow}, {Bartolom{\'e}}, {Bassilana},
  {Bauchet}, {Baudesson-Stella}, {Becciani}, {Bellazzini}, {Bernet}, {Bertone},
  {Bianchi}, {Blanco-Cuaresma}, {Boch}, {Bombrun}, {Bossini}, {Bouquillon},
  {Bragaglia}, {Bramante}, {Breedt}, {Bressan}, {Brouillet}, {Bucciarelli},
  {Burlacu}, {Busonero}, {Butkevich}, {Buzzi}, {Caffau}, {Cancelliere},
  {C{\'a}novas}, {Cantat-Gaudin}, {Carballo}, {Carlucci}, {Carnerero},
  {Carrasco}, {Casamiquela}, {Castellani}, {Castro-Ginard}, {Castro Sampol},
  {Chaoul}, {Charlot}, {Chemin}, {Chiavassa}, {Cioni}, {Comoretto}, {Cooper},
  {Cornez}, {Cowell}, {Crifo}, {Crosta}, {Crowley}, {Dafonte}, {Dapergolas},
  {David}, {David}, {de Laverny}, {De Luise}, {De March}, {De Ridder}, {de
  Souza}, {de Teodoro}, {de Torres}, {del Peloso}, {del Pozo}, {Delbo},
  {Delgado}, {Delgado}, {Delisle}, {Di Matteo}, {Diakite}, {Diener},
  {Distefano}, {Dolding}, {Eappachen}, {Edvardsson}, {Enke}, {Esquej}, {Fabre},
  {Fabrizio}, {Faigler}, {Fedorets}, {Fernique}, {Fienga}, {Figueras},
  {Fouron}, {Fragkoudi}, {Fraile}, {Franke}, {Gai}, {Garabato},
  {Garcia-Gutierrez}, {Garc{\'\i}a-Torres}, {Garofalo}, {Gavras}, {Gerlach},
  {Geyer}, {Giacobbe}, {Gilmore}, {Girona}, {Giuffrida}, {Gomel}, {Gomez},
  {Gonzalez-Santamaria}, {Gonz{\'a}lez-Vidal}, {Granvik},
  {Guti{\'e}rrez-S{\'a}nchez}, {Guy}, {Hauser}, {Haywood}, {Helmi}, {Hidalgo},
  {Hilger}, {H{\l}adczuk}, {Hobbs}, {Holland}, {Huckle}, {Jasniewicz},
  {Jonker}, {Juaristi Campillo}, {Julbe}, {Karbevska}, {Kervella}, {Khanna},
  {Kochoska}, {Kontizas}, {Kordopatis}, {Korn}, {Kostrzewa-Rutkowska},
  {Kruszy{\'n}ska}, {Lambert}, {Lanza}, {Lasne}, {Le Campion}, {Le Fustec},
  {Lebreton}, {Lebzelter}, {Leccia}, {Leclerc}, {Lecoeur-Taibi}, {Liao},
  {Licata}, {Lindstr{\o}m}, {Lister}, {Livanou}, {Lobel}, {Madrero Pardo},
  {Managau}, {Mann}, {Marchant}, {Marconi}, {Marcos Santos}, {Marinoni},
  {Marocco}, {Marshall}, {Martin Polo}, {Mart{\'\i}n-Fleitas}, {Masip},
  {Massari}, {Mastrobuono-Battisti}, {Mazeh}, {McMillan}, {Messina},
  {Michalik}, {Millar}, {Mints}, {Molina}, {Molinaro}, {Moln{\'a}r},
  {Montegriffo}, {Mor}, {Morbidelli}, {Morel}, {Morris}, {Mulone}, {Munoz},
  {Muraveva}, {Murphy}, {Musella}, {Noval}, {Ord{\'e}novic}, {Orr{\`u}},
  {Osinde}, {Pagani}, {Pagano}, {Palaversa}, {Palicio}, {Panahi}, {Pawlak},
  {Pe{\~n}alosa Esteller}, {Penttil{\"a}}, {Piersimoni}, {Pineau}, {Plachy},
  {Plum}, {Poggio}, {Poretti}, {Poujoulet}, {Pr{\v{s}}a}, {Pulone}, {Racero},
  {Ragaini}, {Rainer}, {Raiteri}, {Rambaux}, {Ramos}, {Ramos-Lerate}, {Re
  Fiorentin}, {Regibo}, {Reyl{\'e}}, {Ripepi}, {Riva}, {Rixon}, {Robichon},
  {Robin}, {Roelens}, {Rohrbasser}, {Romero-G{\'o}mez}, {Rowell}, {Royer},
  {Rybicki}, {Sadowski}, {Sagrist{\`a} Sell{\'e}s}, {Sahlmann}, {Salgado},
  {Salguero}, {Samaras}, {Sanchez Gimenez}, {Sanna}, {Santove{\~n}a},
  {Sarasso}, {Schultheis}, {Sciacca}, {Segol}, {Segovia}, {S{\'e}gransan},
  {Semeux}, {Shahaf}, {Siddiqui}, {Siebert}, {Siltala}, {Slezak}, {Smart},
  {Solano}, {Solitro}, {Souami}, {Souchay}, {Spagna}, {Spoto}, {Steele},
  {Steidelm{\"u}ller}, {Stephenson}, {S{\"u}veges}, {Szabados}, {Szegedi-Elek},
  {Taris}, {Tauran}, {Taylor}, {Teixeira}, {Thuillot}, {Tonello}, {Torra},
  {Torra}, {Turon}, {Unger}, {Vaillant}, {van Dillen}, {Vanel}, {Vecchiato},
  {Viala}, {Vicente}, {Voutsinas}, {Weiler}, {Wevers}, {Wyrzykowski}, {Yoldas},
  {Yvard}, {Zhao}, {Zorec}, {Zucker}, {Zurbach}, \& {Zwitter}}]{GaiaEDR3}
{Gaia Collaboration}, {Brown}, A.~G.~A., {Vallenari}, A., {et~al.} 2021, \aap,
  649, A1

\bibitem[{{Gehrz} {et~al.}(1998){Gehrz}, {Truran}, {Williams}, \&
  {Starrfield}}]{Gehrz98}
{Gehrz}, R.~D., {Truran}, J.~W., {Williams}, R.~E., \& {Starrfield}, S. 1998,
  \pasp, 110, 3

\bibitem[{{Gill} \& {O'Brien}(1998)}]{GillOBrien98}
{Gill}, C.~D. \& {O'Brien}, T.~J. 1998, \mnras, 300, 221

\bibitem[{{Haefner} \& {Metz}(1982)}]{Haefner&Metz1982}
{Haefner}, R. \& {Metz}, K. 1982, \aap, 109, 171

\bibitem[{{Harvey} {et~al.}(2016){Harvey}, {Redman}, {Boumis}, \&
  {Akras}}]{Harvey16}
{Harvey}, E., {Redman}, M.~P., {Boumis}, P., \& {Akras}, S. 2016, \aap, 595,
  A64

\bibitem[{{Hillman} {et~al.}(2020){Hillman}, {Shara}, {Prialnik}, \&
  {Kovetz}}]{Hillman20}
{Hillman}, Y., {Shara}, M.~M., {Prialnik}, D., \& {Kovetz}, A. 2020, Nature
  Astronomy, 4, 886

\bibitem[{{I{\l}kiewicz} {et~al.}(2021){I{\l}kiewicz}, {Scaringi}, {Court},
  {Maccarone}, {Altamirano}, {Bradshaw}, {Degenaar}, {Fratta}, {Littlefield},
  {Shahbaz}, \& {Wijnands}}]{Ilkiewicz21_AQMen}
{I{\l}kiewicz}, K., {Scaringi}, S., {Court}, J. M.~C., {et~al.} 2021, \mnras,
  503, 4050

\bibitem[{{Kubiak}(1984)}]{Kubiak1984}
{Kubiak}, M. 1984, \actaa, 34, 331

\bibitem[{{Lasota} {et~al.}(1999){Lasota}, {Kuulkers}, \& {Charles}}]{Lasota99}
{Lasota}, J.-P., {Kuulkers}, E., \& {Charles}, P. 1999, \mnras, 305, 473

\bibitem[{{Liimets} {et~al.}(2012){Liimets}, {Corradi},
  {Santander-Garc{\'\i}a}, {Villaver}, {Rodr{\'\i}guez-Gil}, {Verro}, \&
  {Kolka}}]{Liimets12}
{Liimets}, T., {Corradi}, R.~L.~M., {Santander-Garc{\'\i}a}, M., {et~al.} 2012,
  \apj, 761, 34

\bibitem[{{Lopes de Oliveira} {et~al.}(2020){Lopes de Oliveira}, {Bruch},
  {Rodrigues}, {Oliveira}, \& {Mukai}}]{LopezdeOliveira20_CTCVJ2056-3014}
{Lopes de Oliveira}, R., {Bruch}, A., {Rodrigues}, C.~V., {Oliveira}, A.~S., \&
  {Mukai}, K. 2020, \apjl, 898, L40

\bibitem[{{Lunt}(1926)}]{Lunt1926}
{Lunt}, J. 1926, \mnras, 86, 498

\bibitem[{{Luri} {et~al.}(2018){Luri}, {Brown}, {Sarro}, {Arenou},
  {Bailer-Jones}, {Castro-Ginard}, {de Bruijne}, {Prusti}, {Babusiaux}, \&
  {Delgado}}]{Luri18}
{Luri}, X., {Brown}, A.~G.~A., {Sarro}, L.~M., {et~al.} 2018, \aap, 616, A9

\bibitem[{{Luridiana} {et~al.}(2015){Luridiana}, {Morisset}, \&
  {Shaw}}]{Luridiana+2015_pyneb}
{Luridiana}, V., {Morisset}, C., \& {Shaw}, R.~A. 2015, \aap, 573, A42

\bibitem[{{Lyke} \& {Campbell}(2009)}]{Lyke09}
{Lyke}, J.~E. \& {Campbell}, R.~D. 2009, \aj, 138, 1090

\bibitem[{{Macfarlane} {et~al.}(2014){Macfarlane}, {Steeghs}, \&
  {Woudt}}]{Macfarlane14}
{Macfarlane}, S., {Steeghs}, D., \& {Woudt}, P. 2014, in Astronomical Society
  of the Pacific Conference Series, Vol. 490, Stellar Novae: Past and Future
  Decades, ed. P.~A. {Woudt} \& V.~A.~R.~M. {Ribeiro}, 115

\bibitem[{{Moraes} \& {Diaz}(2009)}]{Moraes09}
{Moraes}, M. \& {Diaz}, M. 2009, \aj, 138, 1541

\bibitem[{{Nelemans} {et~al.}(2016){Nelemans}, {Siess}, {Repetto}, {Toonen}, \&
  {Phinney}}]{Nelemans+16}
{Nelemans}, G., {Siess}, L., {Repetto}, S., {Toonen}, S., \& {Phinney}, E.~S.
  2016, \apj, 817, 69

\bibitem[{{Onken} {et~al.}(2019){Onken}, {Wolf}, {Bessell}, {Chang}, {Da
  Costa}, {Luvaul}, {Mackey}, {Schmidt}, \& {Shao}}]{SkyMapperDR2}
{Onken}, C.~A., {Wolf}, C., {Bessell}, M.~S., {et~al.} 2019, \pasa, 36, e033

\bibitem[{{Osterbrock} \& {Ferland}(2006)}]{Osterbrock2006}
{Osterbrock}, D.~E. \& {Ferland}, G.~J. 2006, {Astrophysics of gaseous nebulae
  and active galactic nuclei}

\bibitem[{{Porter} {et~al.}(1998){Porter}, {O'Brien}, \& {Bode}}]{Porter98}
{Porter}, J.~M., {O'Brien}, T.~J., \& {Bode}, M.~F. 1998, \mnras, 296, 943

\bibitem[{{Ramsay} {et~al.}(2017){Ramsay}, {Schreiber}, {G{\"a}nsicke}, \&
  {Wheatley}}]{Ramsay17}
{Ramsay}, G., {Schreiber}, M.~R., {G{\"a}nsicke}, B.~T., \& {Wheatley}, P.~J.
  2017, \aap, 604, A107

\bibitem[{{Reynolds}(2008)}]{Reynolds08}
{Reynolds}, S.~P. 2008, \araa, 46, 89

\bibitem[{{Ribeiro} \& {Diaz}(2006)}]{Ribeiro06}
{Ribeiro}, F. M.~A. \& {Diaz}, M.~P. 2006, \pasp, 118, 84

\bibitem[{{Rodrigo} \& {Solano}(2020)}]{SVOFilterService}
{Rodrigo}, C. \& {Solano}, E. 2020, in XIV.0 Scientific Meeting (virtual) of
  the Spanish Astronomical Society, 182

\bibitem[{{Sahman} {et~al.}(2018){Sahman}, {Dhillon}, {Littlefair}, \&
  {Hallinan}}]{Sahman18}
{Sahman}, D.~I., {Dhillon}, V.~S., {Littlefair}, S.~P., \& {Hallinan}, G. 2018,
  \mnras, 477, 4483

\bibitem[{{Santamar{\'\i}a} {et~al.}(2019){Santamar{\'\i}a}, {Guerrero},
  {Ramos-Larios}, {Sabin}, {V{\'a}zquez}, {G{\'o}mez-Mu{\~n}oz}, \&
  {Toal{\'a}}}]{Santamaria19}
{Santamar{\'\i}a}, E., {Guerrero}, M.~A., {Ramos-Larios}, G., {et~al.} 2019,
  \mnras, 483, 3773

\bibitem[{{Santamar{\'\i}a} {et~al.}(2020){Santamar{\'\i}a}, {Guerrero},
  {Ramos-Larios}, {Toal{\'a}}, {Sabin}, {Rubio}, \&
  {Quino-Mendoza}}]{Santamaria20}
{Santamar{\'\i}a}, E., {Guerrero}, M.~A., {Ramos-Larios}, G., {et~al.} 2020,
  \apj, 892, 60

\bibitem[{{Santamar{\'\i}a} {et~al.}(2022{\natexlab{a}}){Santamar{\'\i}a},
  {Guerrero}, {Toal{\'a}}, {Ramos-Larios}, \& {Sabin}}]{Santamaria22b}
{Santamar{\'\i}a}, E., {Guerrero}, M.~A., {Toal{\'a}}, J.~A., {Ramos-Larios},
  G., \& {Sabin}, L. 2022{\natexlab{a}}, \mnras, 517, 2567

\bibitem[{{Santamar{\'\i}a} {et~al.}(2022{\natexlab{b}}){Santamar{\'\i}a},
  {Guerrero}, {Zavala}, {Ramos-Larios}, {Toal{\'a}}, \&
  {Sabin}}]{Santamaria22a}
{Santamar{\'\i}a}, E., {Guerrero}, M.~A., {Zavala}, S., {et~al.}
  2022{\natexlab{b}}, \mnras, 512, 2003

\bibitem[{{Schaefer} {et~al.}(2019){Schaefer}, {Boyd}, {Clayton}, {Frank},
  {Johnson}, {Kemp}, {Pagnotta}, {Patterson}, {Rodr{\'\i}guez Marco}, \&
  {Xiao}}]{Schaefer+19_QZAur}
{Schaefer}, B.~E., {Boyd}, D., {Clayton}, G.~C., {et~al.} 2019, \mnras, 487,
  1120

\bibitem[{{Schmidtobreick} {et~al.}(2008){Schmidtobreick}, {Papadaki},
  {Tappert}, \& {Ederoclite}}]{Schmidtobreick08}
{Schmidtobreick}, L., {Papadaki}, C., {Tappert}, C., \& {Ederoclite}, A. 2008,
  \mnras, 389, 1345

\bibitem[{{Schmidtobreick} {et~al.}(2015){Schmidtobreick}, {Shara}, {Tappert},
  {Bayo}, \& {Ederoclite}}]{Schmidtobreick15}
{Schmidtobreick}, L., {Shara}, M., {Tappert}, C., {Bayo}, A., \& {Ederoclite},
  A. 2015, \mnras, 449, 2215

\bibitem[{{Schmidtobreick} {et~al.}(2003){Schmidtobreick}, {Tappert}, \&
  {Saviane}}]{Schmidtobreick+2003}
{Schmidtobreick}, L., {Tappert}, C., \& {Saviane}, I. 2003, \mnras, 342, 145

\bibitem[{{Shara} {et~al.}(2017){Shara}, {Drissen}, {Martin}, {Alarie}, \&
  {Stephenson}}]{Shara+17_ATCnc}
{Shara}, M.~M., {Drissen}, L., {Martin}, T., {Alarie}, A., \& {Stephenson},
  F.~R. 2017, \mnras, 465, 739

\bibitem[{{Shore} {et~al.}(2018){Shore}, {Kuin}, {Mason}, \& {De Gennaro
  Aquino}}]{Shore+2018_dust}
{Shore}, S.~N., {Kuin}, N.~P., {Mason}, E., \& {De Gennaro Aquino}, I. 2018,
  \aap, 619, A104

\bibitem[{{Sion}(1999)}]{Sion99}
{Sion}, E.~M. 1999, \pasp, 111, 532

\bibitem[{{Sion} {et~al.}(2017){Sion}, {Godon}, \& {Jones}}]{Sion17}
{Sion}, E.~M., {Godon}, P., \& {Jones}, L. 2017, \aj, 153, 109

\bibitem[{{Slavin} {et~al.}(1995){Slavin}, {O'Brien}, \& {Dunlop}}]{Slavin95}
{Slavin}, A.~J., {O'Brien}, T.~J., \& {Dunlop}, J.~S. 1995, \mnras, 276, 353

\bibitem[{{Spencer Jones}(1931)}]{SpencerJones31}
{Spencer Jones}, H. 1931, Annals of the Cape Observatory, 10, 9.1

\bibitem[{{Starrfield} {et~al.}(2016){Starrfield}, {Iliadis}, \&
  {Hix}}]{Starrfield16}
{Starrfield}, S., {Iliadis}, C., \& {Hix}, W.~R. 2016, \pasp, 128, 051001

\bibitem[{{Storey} \& {Hummer}(1995)}]{Storey&Hummer1995}
{Storey}, P.~J. \& {Hummer}, D.~G. 1995, \mnras, 272, 41

\bibitem[{{Strope} {et~al.}(2010){Strope}, {Schaefer}, \& {Henden}}]{Strope10}
{Strope}, R.~J., {Schaefer}, B.~E., \& {Henden}, A.~A. 2010, \aj, 140, 34

\bibitem[{{Takeda} {et~al.}(2022){Takeda}, {Diaz}, {Campbell}, {Lyke},
  {Lawrence}, {Linford}, \& {Sokolovsky}}]{Takeda22}
{Takeda}, L., {Diaz}, M., {Campbell}, R.~D., {et~al.} 2022, \mnras, 511, 1591

\bibitem[{{Takei} {et~al.}(2015){Takei}, {Drake}, {Yamaguchi}, {Slane},
  {Uchiyama}, \& {Katsuda}}]{Takei+15}
{Takei}, D., {Drake}, J.~J., {Yamaguchi}, H., {et~al.} 2015, \apj, 801, 92

\bibitem[{{Tappert} {et~al.}(2023){Tappert}, {Celed{\'o}n}, \&
  {Schmidtobreick}}]{Tappert23}
{Tappert}, C., {Celed{\'o}n}, L., \& {Schmidtobreick}, L. 2023, arXiv e-prints,
  arXiv:2309.03777

\bibitem[{{Tappert} {et~al.}(2020){Tappert}, {Vogt}, {Ederoclite},
  {Schmidtobreick}, {Vu{\v{c}}kovi{\'c}}, \& {Becegato}}]{Tappert20}
{Tappert}, C., {Vogt}, N., {Ederoclite}, A., {et~al.} 2020, \aap, 641, A122

\bibitem[{{Vogt} {et~al.}(2017){Vogt}, {Schreiber}, {Hambsch}, {Retamales},
  {Tappert}, {Schmidtobreick}, \& {Fuentes-Morales}}]{Vogt+2017}
{Vogt}, N., {Schreiber}, M.~R., {Hambsch}, F.~J., {et~al.} 2017, \pasp, 129,
  014201

\bibitem[{{Weilbacher} {et~al.}(2020){Weilbacher}, {Palsa}, {Streicher},
  {Bacon}, {Urrutia}, {Wisotzki}, {Conseil}, {Husemann}, {Jarno}, {Kelz},
  {P{\'e}contal-Rousset}, {Richard}, {Roth}, {Selman}, \&
  {Vernet}}]{Weilbacher20}
{Weilbacher}, P.~M., {Palsa}, R., {Streicher}, O., {et~al.} 2020, \aap, 641,
  A28

\bibitem[{{Williams} \& {Gallagher}(1979)}]{Williams79}
{Williams}, R.~E. \& {Gallagher}, J.~S. 1979, \apj, 228, 482

\bibitem[{{Wolf} {et~al.}(2018){Wolf}, {Onken}, {Luvaul}, {Schmidt}, {Bessell},
  {Chang}, {Da Costa}, {Mackey}, {Martin-Jones}, {Murphy}, {Preston}, {Scalzo},
  {Shao}, {Smillie}, {Tisserand}, {White}, \& {Yuan}}]{SkyMapperDR1}
{Wolf}, C., {Onken}, C.~A., {Luvaul}, L.~C., {et~al.} 2018, \pasa, 35, e010

\bibitem[{{Woudt} {et~al.}(2009){Woudt}, {Steeghs}, {Karovska}, {Warner},
  {Groot}, {Nelemans}, {Roelofs}, {Marsh}, {Nagayama}, {Smits}, \&
  {O'Brien}}]{Woudt09}
{Woudt}, P.~A., {Steeghs}, D., {Karovska}, M., {et~al.} 2009, \apj, 706, 738

\bibitem[{{Wright} {et~al.}(2010){Wright}, {Eisenhardt}, {Mainzer}, {Ressler},
  {Cutri}, {Jarrett}, {Kirkpatrick}, {Padgett}, {McMillan}, {Skrutskie},
  {Stanford}, {Cohen}, {Walker}, {Mather}, {Leisawitz}, {Gautier}, {McLean},
  {Benford}, {Lonsdale}, {Blain}, {Mendez}, {Irace}, {Duval}, {Liu}, {Royer},
  {Heinrichsen}, {Howard}, {Shannon}, {Kendall}, {Walsh}, {Larsen}, {Cardon},
  {Schick}, {Schwalm}, {Abid}, {Fabinsky}, {Naes}, \& {Tsai}}]{WISE2010}
{Wright}, E.~L., {Eisenhardt}, P. R.~M., {Mainzer}, A.~K., {et~al.} 2010, \aj,
  140, 1868

\end{thebibliography}

\end{document}